\documentclass[11pt]{revtex4}
\usepackage{blindtext}
\usepackage{amsfonts}
\usepackage{graphicx}
\usepackage{epsfig}
\usepackage{float}
\usepackage{bm}
\usepackage{epstopdf}
\usepackage{url}
\usepackage{hyperref}
\usepackage{xcolor}
\usepackage{amsmath}
\usepackage{amssymb}

\begin{document}

\title{2D Skyrmion topological charge of spin textures with arbitrary boundary conditions: a two-component spinorial BEC as a case study}
\author{S. S\'anchez-Res\'endiz, E. Neri, S. Gonz\'alez-Hern\'andez, and V. Romero-Roch\'{\i}n}
\email{romero@fisica.unam.mx}
\address{\it Instituto de F\'{\i}sica, Universidad
Nacional Aut\'onoma de M\'exico, Apartado Postal 20-364, 01000 Ciudad de M\'exico, Mexico.}

\begin{abstract}

We derive the most general expression for the Skyrmion topological charge for a two-dimensional spin texture, valid for any type of boundary conditions or for any arbitrary spatial region within the texture. It reduces to the usual one $Q = 1/4\pi \iint \vec f \cdot \left(\partial_x \vec f \times \partial_y \vec f\right)$ for the appropriate boundary conditions, with $\vec f$ the spin texture. The general expression resembles the Gauss-Bonet theorem for the Euler-Poincar\'e characteristic of a 2D surface, but it has definite differences, responsible for the assignment of the proper signs of the Skyrmion charges. Additionally, we show that the charge of a single Skyrmion is the product of the value of the  normal component of the spin texture at the singularity times the Index or winding number of the transverse texture, a generalization of a Poincar\'e theorem. We illustrate our general results analyzing in detail a two-component spinor Bose-Einstein condensate in 2D in the presence of an external magnetic field, via the Gross-Pitaevskii equation. The condensate spin textures present Skyrmions singularities at the spatial locations where the transverse magnetic field vanishes. We show that the ensuing superfluid vortices and Skyrmions have the same value for their corresponding topological charges, in turn due to the structure of the magnetic field.
\end{abstract}
\maketitle

\medskip

\section{Introduction}

Skyrmions are topological defects or excitations that emerge in continuous fields and spin textures due to intrinsic chiral interactions  or to the presence of external inhomogeneous magnetic or gauge fields. These excitations arise in diverse materials or fields as in nuclear physics \cite{Skyrme,Witten,Meissner}, particle physics \cite{Holzwarth,Schwesinger}, notably in magnetic systems in condensed matter, such as in 2D electron gas \cite{Brey}, magnetic metals \cite{Rossler,Hoffmann,Romming}, transition metal films \cite{Dupe}, frustrated magnets \cite{Leonov} and in a host of other magnetic situations \cite{Nagaosa,Jalil}, liquid crystals \cite{Wright,Leonov-Cris} and, the context of the present paper, in Bose-Einstein condensates, both 2D \cite{Stoof,Kasamatsu,Leanhardt,Pietila,Leslie,Choi-PRL,Su,Xu,Liu,Huang,Choi,Hu,Lovegrove,Borgh,Dong,Roberto,Liu-2024} and 3D  \cite{Marzlin,Ruostekoski,Battye,Mizushima,Pietila-3D,Kawakami,Liu-3D,Luo-3D,Chen-3D,Tiurev}. Ever since Skyrme seminal work \cite{Skyrme}, its analysis has been pursued in field theoretical models such as the non-linear $\sigma$-model in high energy physics \cite{Witten,Meissner,Wilczek-Zee}, or in terms of the Dzyaloshinskii-Moriya theory in magnetic systems \cite{Dzyaloshinsky,Moriya}, or within the Gross-Pitaevskii (GP) equation in Bose-Einstein condensates, as in essentially all of the theoretical works already referenced. In this paper we deal with Skyrmions generated by an external gauge field in a two-component spinorial Bose-Einstein condensate in two spatial dimensions, described by two coupled GP equations. This stylized model allows for a simple and clear description of  spin textures that can show a rich variety of Skyrmions, with the allowance for the determination of all its  inherent and topological properties. In most of the experimental Skyrmion realizations, BEC studies have been performed in real alkaline weakly interacting Bose clouds with spin $F=1$  \cite{Leanhardt,Choi,Choi-PRL,Su} and  $F = 2$ \cite{Leslie}, however, the topological properties of a given spin structure should be independent of their physical origin.\\

\begin{figure}[htbp]
\begin{center}
\includegraphics[scale=0.3]{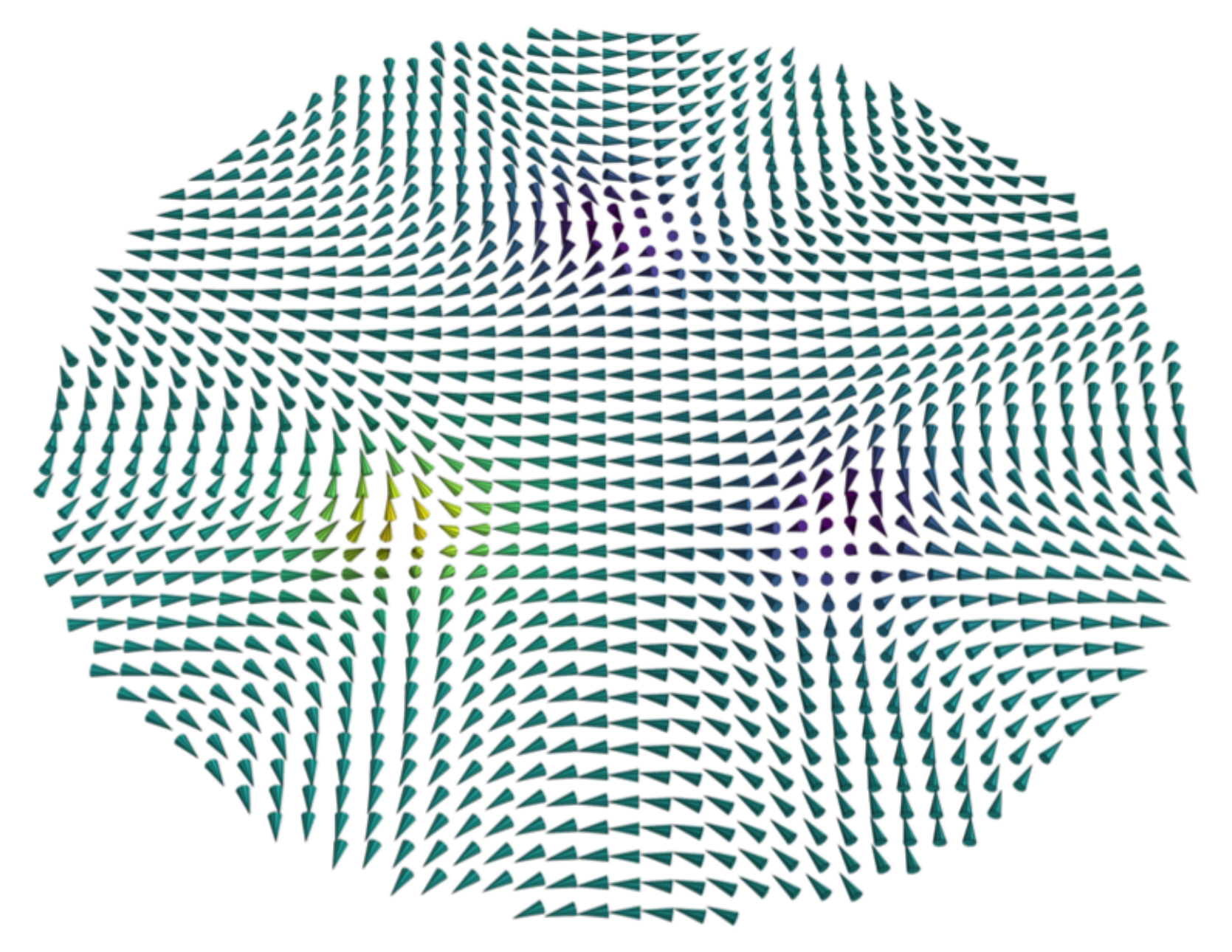}
\end{center}
\caption{A spin texture $\vec f(x,y)$ showing three Skyrmions, arising from the solution of the GP equation of a two-component spinorial 2D BEC in the presence of a magnetic field that vanishes at the locations where the texture becomes $\vec f = \pm z$. See Appendix for calculation details.
} \label{Skyrm-2-1-A}
\end{figure}

\begin{figure}[htbp]
\begin{center}
\includegraphics[scale=0.3]{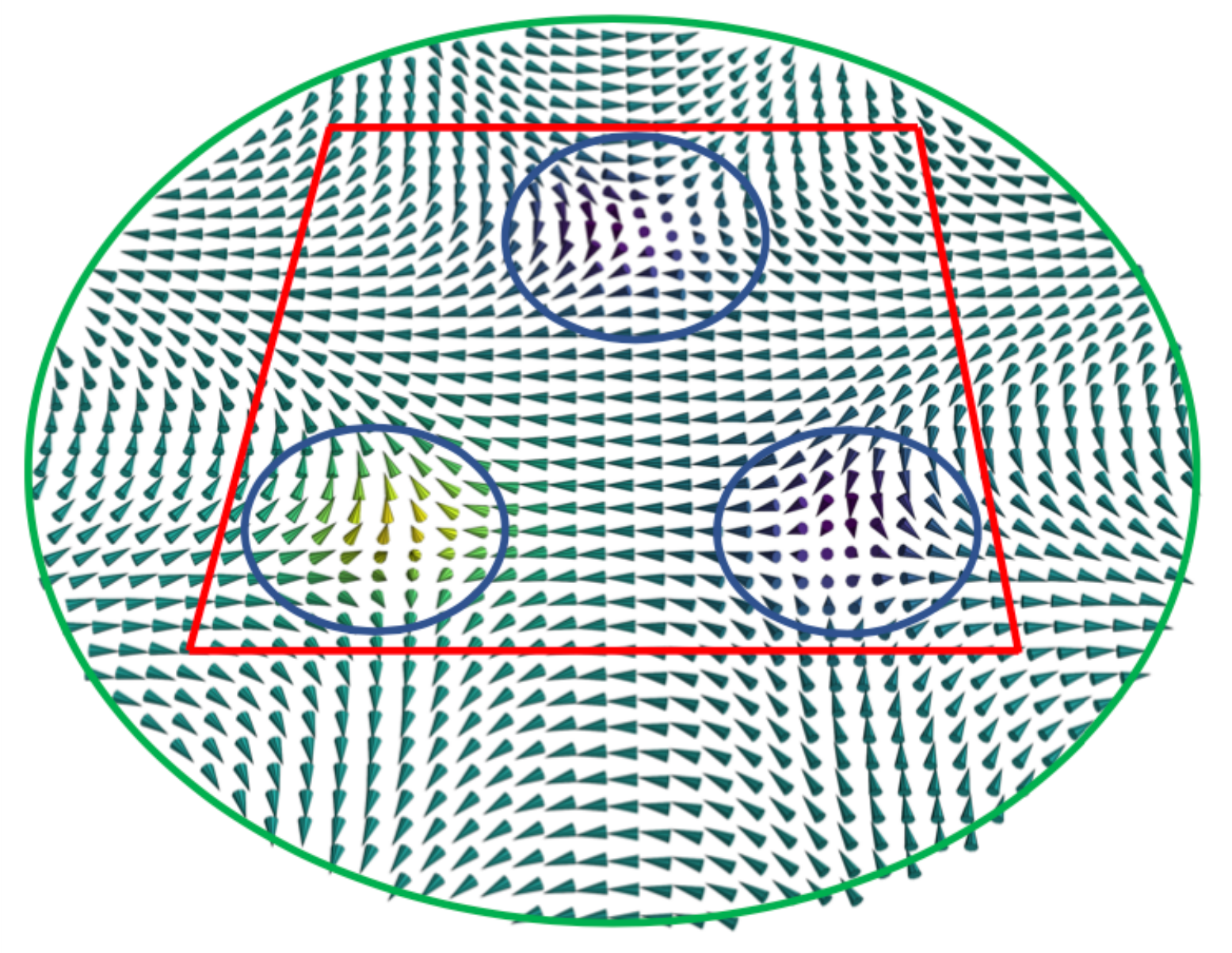}
\end{center}
\caption{The same spin texture as in figure \ref{Skyrm-2-1-A}. The contours in color represent several boundary curves $\vec r_c(t)$ within which the Skyrmion charge $Q_S$ can be calculated using (\ref{QS-final}). The small circles (blue) enclose the local Skyrmions yielding $Q_S = + 1$ for the two on the right and $Q_S = -1$ for the other. The square (red) and the large circle (green) enclose the full configuration yielding $Q_S = +1$.
} \label{Skyrm-2-1-B}
\end{figure}

An outstanding feature of Skyrmions is their topological character ingrained and described by its topological charge. In most of the studies on 2D Skyrmions  the topological charge has been considered to be given by the expression \cite{Wilczek-Zee,Kasamatsu,Heinze,Romming,Choi,Hu,Lovegrove,Dong,Choi-PRL,Pietila,Dupe,Shnir}
\begin{equation}
Q = \frac{1}{4\pi} \iint \vec f \cdot \left( \frac{\partial \vec f}{\partial x} \times \frac{\partial \vec f}{\partial y} \right) \>dxdy \>, \label{Q}
\end{equation}
where $\vec f(x,y)$ is the 3D vector spin texture or continuous field of the problem at hand, obeying $\vec f \cdot \vec f = 1$ everywhere in the $XY$-plane. It is expected that the charge $Q$ is an integer. This expression, however, requires particular boundary conditions. Skyrme \cite{Skyrme} and others afterwards \cite{Wilczek-Zee,Kasamatsu,Shnir}  considered that $\vec f =$ constant as $|\vec r| \to \infty$ and, in such a case, $Q$ is a positive or negative multiple integer, with the interpretation that $Q$ is the number of times the mapping of $\vec f$ covers the unit sphere. Additionally, if the spin texture lies on the plane at the boundary, then $Q = \pm 1/2$, a defect called a {\it half} Skyrmion \cite{Su,Leslie,Choi-PRL,Liu,Roberto}. These boundary conditions, however, need not appear or exist in a given physical problem \cite{Roberto} or in a multiple Skyrmion configuration, where one may need to evaluate the local charge of the Skyrmions present. In these cases expression (\ref{Q}) cannot account for the calculation of the whole charge that, from topological consideration, must be an integer. As an example, in figure \ref{Skyrm-2-1-A} we present a spin texture arising in the 2D spin 1/2 spinor Bose-Einstein condensate (BEC) here studied, the details of such a solution given further below. We observe a configuration of three Skyrmions, one pointing on the positive $z$-direction and the other two in the negative one. Each defect should have its own topological charge, properly defined, and the total one of the full configuration should be the sum of the individual ones. The main purpose of this article is to show that a topological charge $Q_S$ of the whole or part of the spin texture configuration can be defined within an {\it arbitrary} closed curve, which may or may not be the true boundary of the spin texture. Parametrizing the curve by $\vec r_c(t)$, with $t: 0 \to T$ and  $\vec r_c(0)= \vec r_c(T)$, we will find that such a topological charge is given by  
\begin{eqnarray}
2\pi Q_S &=& \iint_A \> \vec f \cdot \left(\frac{\partial \vec f}{\partial x} \times \frac{\partial \vec f}{\partial x} \right) \>dx dy + \oint_{\vec r_c} \vec f \cdot 
\frac{\left( \frac{d \vec f}{dt} \times \frac{d^2\vec f}{dt^2} \right) }{ \left|\frac{d \vec f}{dt}\right|^2 }\
dt  \nonumber \\
&& + \sum_{i=1}^s \sin^{-1} \vec f_i \cdot \frac{\left( \left. \frac{d \vec f}{dt}\right|_i^{\> +} \times \left. \frac{d \vec f}{dt}\right|_i^{\> -}\right)}{\left| \frac{d \vec f}{dt}\right|_i^{\> +}\left| \frac{d \vec f}{dt}\right|_i^{\> -} }\>, \label{QS-final}
\end{eqnarray}
with $Q_S \in \mathbb{Z}$, a positive or negative integer. The first term is the usual one, except for a factor of 2, with $A$ the area enclosed  by the curve $\vec r_c$. The second is a boundary term with the integral along the curve $\vec r_c$, and the third one takes into account the possible fact that the vector $\vec t = (d \vec f/dt)/|d \vec f/dt|$, tangent to the parametric 3D curve $\vec f(\vec r_c(t))$, 
may be discontinuous at a finite number of $s$ points along the curve. All these aspects should become clearer in the derivation given below. The above expression indeed reduces to the usual one $Q_S = 2 Q$ for the boundary conditions described in the previous paragraph. Figure \ref{Skyrm-2-1-B} shows the same spin texture as in figure \ref{Skyrm-2-1-A} but with several boundary curves $\vec r_c$. The small circular ones each enclose the local Skyrmion singularities, yielding $Q_S = +1$ for the  two of them pointing down and $Q_S = -1$ for the one pointing up, using (\ref{QS-final}). The large circle and the square curve enclose the three Skyrmions and yield $Q_S = +1$, which is the sum of the local ones. We remark that, for the square curve, the last term of (\ref{QS-final}) must be taken into account since it gives rise to discontinuities on the tangent vector mentioned above.\\

While we claim that the above expression is a general one for any spin texture, since the derivation that we provide makes no reference to the physical origin of the spin texture, for purposes of illustrations and for a better understanding of the Skyrmion charge, we will use the spin textures arising from  a spinorial BEC in the presence of an external magnetic field, as a case study. In addition, an important and interesting point, is that such a system shows another well known type of topological defects, namely, quantum vortices \cite{Kasamatsu,Lin,Kawaguchi,gallery,Stamper,Ueda}. We will explicitly show the profound relationship between the quantum vortices and the Skyrmions, namely, that their corresponding local charges have exactly the same values. The present stylized model allows for finding that, for the present problem, the external magnetic or gauge field ultimately defines the topological structure of the vortices and the Skyrmions.\\

The article is organized as follows. In Sections II and III we present a detailed analysis, both in a simplified analytic and in a full numerical fashion, of stationary solutions to the Gross-Pitaevskii equations of a 2D 1/2 spinorial Bose gas in the presence of an external inhomogeneous magnetic field. We show how the quantum vortices and Skyrmions can be predicted from the knowledge of the magnetic field. In Section IV, the main contribution of this article, we provide a general derivation of the topological charge of Skyrmions present in any given spin texture, for arbitrary boundary curves. Section V is devoted to a discussion between the similarities and discrepancies between the Skyrmions topological charge and the Euler-Poincar\'e characteristic of a surface in terms of the Gauss-Bonet theorem \cite{Weatherburn,Kreyszig,McCleary,doCarmo}. 
In an Appendix we illustrate our main result with a variety of numerical solutions of different spin textures generated by different magnetic fields, and for different types of arbitrary boundary curves. Final remarks are given in Section VI.

\section{1/2 Spinor BEC in the presence of an external magnetic field}

We consider stationary states of a two-component spinorial BEC in 2D, with a Zeeman coupling to an external magnetic field, as described by the coupled Gross-Pitaevskii (GP) equations
\begin{equation}
\mu\psi_\alpha(\vec r) = \left[-\frac{\hbar^2}{2m}\nabla^2 + \frac{1}{2} m\omega^2 \vec r^{\> 2}
+ g (|\psi_+|^2 + |\psi_-|^2) \right]\psi_\alpha(\vec r) - \vec B(\vec r)\cdot\vec \mu^{\>\alpha\beta} \psi_\beta(\vec r) \>,\label{GP}
\end{equation}
where $\vec r = (x,y)$ and $\alpha = +,-$ denotes the two components of the spinor. Summing over repeated indices is assumed. $\mu$ is the chemical potential and the magnetic dipole operator is given by
\begin{equation}
\vec \mu^{\>\alpha\beta} = \mu_0 \vec \sigma^{\>\alpha\beta} \>,\label{mu}
\end{equation}
where $\vec \sigma = \hat x \sigma_x + \hat y \sigma_y + \hat z \sigma_z$  are Pauli matrices and  $\mu_0 > 0$ the magnitude of the magnetic dipole moment. The magnetic field is 
\begin{equation}
\vec B(\vec r) = \hat x B_x(\vec r) + \hat y B_y(\vec r) + \hat z B_z \>.
\end{equation}
We denote $\vec B_T(\vec r) = B_x(\vec r) \hat x + B_y(\vec r) \hat y$ as the {\it transverse} magnetic field and, for simplicity, we will consider $B_z(\vec r) =$ constant . We choose  $\vec B_T(\vec r)$  such that it vanishes at certain points $\vec r_1, \vec r_2, \dots, \vec r_s$. We shall see that at those points there appear quantum vortices and the origins of Skyrmions \cite{gallery}.  The normalization condition is
\begin{equation}
\int d^2 r \> (|\psi_1|^2 + |\psi_2|^2) = 1 \>.
\end{equation}

Writing the solution as $\psi_\alpha = \sqrt{\rho_\alpha} \> e^{i \Theta_\alpha}$, the superfluid density $\rho_\alpha$ and  the superfluid velocity $\vec v_\alpha$, in each component, are given by
\begin{eqnarray}
\rho_\alpha &=& |\psi_\alpha|^2 \label{rho} \>,\\
\vec v_\alpha & = & \frac{\hbar}{m} \nabla \Theta_\alpha \>.\label{vel}
\end{eqnarray}

The circulation $C$ of the velocity field in an arbitrary closed circuit ${\cal C}$ is defined by
\begin{equation}
C =  \oint_{\cal C} \vec v_\alpha \cdot d\vec r \>.\label{circ}
\end{equation}
As we will explicitly calculate, if the circuit does not enclose a quantum vortex the circulation vanishes, but when a vortex or vortices are enclosed, the circulation yields the vortex topological charge $Q_V$ within the circuit \cite{Ueda,gallery}
\begin{equation}
C =  \frac{2 \pi \hbar}{m} Q_V\>
\end{equation}
with $Q_V \in \mathbb{Z}$, a positive or negative integer. \\

The condensate spin texture $\vec f(x,y)$ is an observable, independent of the spin basis, and defined at any spatial point within the condensate,
\begin{equation}
\vec f = \frac{1}{\rho_+ + \rho_-}\left(\psi_+^* \>\> \psi_-^*\right) \left[\hat x \sigma_x + \hat y \sigma_y + \hat z \sigma_z\right]
\left(
\begin{array}{c}
\psi_+ \\ \psi_-
\end{array}\right) \>.
\end{equation}
It can be verified that the spin texture has magnitude one $\vec f \cdot \vec f = 1$ everywhere.  \\

Writing $\vec f = \hat x f_x + \hat y f_y + \hat z f_z$, the components explicitly are
\begin{eqnarray}
f_x &=& \frac{\psi_+^* \psi_- + \psi_-^* \psi_+}{\rho_+ + \rho_-}  \nonumber \\
&=& 2 \frac{\sqrt{\rho_+\rho_-}}{\rho_+ + \rho_-} \cos(\Theta_- - \Theta_+) \nonumber \\
f_y &=&- i\frac{\psi_+^* \psi_- - \psi_-^* \psi_+ }{\rho_+ + \rho_-} \nonumber \\
&=& 2\frac{\sqrt{\rho_+\rho_-}}{\rho_+ + \rho_-} \sin(\Theta_- - \Theta_+)\nonumber\\
f_z &=& \frac{\rho_+ - \rho_- }{\rho_+ + \rho_-} \label{f}\>.
\end{eqnarray}

\section{Vortices and Skyrmions in the vicinity of points of vanishing transverse magnetic field}

\begin{figure}[htbp]
\begin{center}
\includegraphics[scale=0.5]{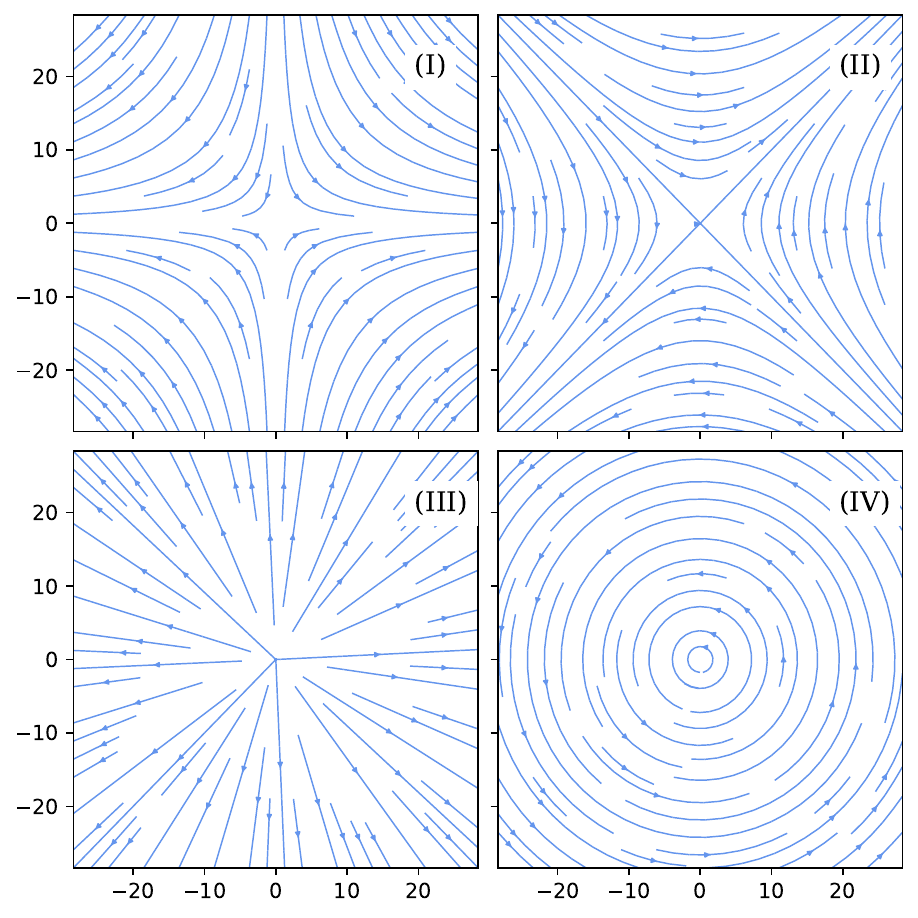}
\end{center}
\caption{Transverse magnetic field lines $\vec B_T = \hat i B_x + \hat j B_y$ on the $XY$-plane. The labels (I)-(IV) correspond to the field types given by (\ref{campos}) for $n = 1$. The field becomes zero at the origin.
} \label{fieldB}
\end{figure}

While we will consider a variety of magnetic fields, all have in common that they have spatial points $\vec r_1, \vec r_2, \dots, \vec r_s$ where the transverse magnetic field vanishes, $\vec B_T (\vec r_i) = 0$. This property imprints both the appearance of vortices and of Skyrmions at those points. To understand how this occurs, with no loss of generality, we consider a transverse magnetic field that vanishes at the origin $\vec r = 0$ and analyze the behavior of the spinor wavefunctions in its vicinity. For this we explicitly write both GP equations for the two spinor components, assuming that $B_z = 0$,
\begin{eqnarray}
\mu\psi_+ &=& \left[-\frac{\hbar^2}{2m}\nabla^2 + \frac{1}{2} m\omega^2 r^2
+ g \rho \right]\psi_+ - \mu_0 (B_x - i B_y) \psi_- \nonumber \\
\mu\psi_- &=& \left[-\frac{\hbar^2}{2m}\nabla^2 + \frac{1}{2} m\omega^2 r^2
+ g \rho \right]\psi_- - \mu_0 (B_x + i B_y) \psi_+ \>,\label{GP2}
\end{eqnarray}
where $\rho = |\psi_+|^2 +  |\psi_+|^2$ and $r= \sqrt{x^2+y^2}$. In the vicinity of the origin the magnetic field generically vanishes as \cite{gallery}
\begin{equation}
B_x - i B_y \approx {\cal B}_0 \> r^n \> e^{i n \varphi} \label{fieldsn}
\end{equation}
where ${\cal B}_0$ is the field strength, not necessarily real, $\tan \varphi = y/x$ and $n$ a positive integer.  We note that if ${\cal B}_0$ is real, the resulting field obeys $\nabla \cdot \vec B = 0$ and $\nabla \times \vec B = 0$, namely, it would be a true magnetic field. However, one can consider arbitrary ``artificial'' gauge fields \cite{Lin-Synthetic,Dalibard-Synthetic} that do not necessarily satisfy Maxwell equations but that allow for different sorts of fields and which in turn give rise to different spin textures, as we show in the rest of the paper. In this way, generically, there are four different types of fields in the vicinity of a vanishing location, with ${\cal B}_0$ real, positive or negative,
\begin{eqnarray} 
{\rm I)} \>\>\>\>B_x - i B_y &\approx& {\cal B}_0 \> r^n \> e^{i n \varphi} \>\>\>\>\>\>\>\>\>\>\>
{\rm II)}\>\>\> B_x - i B_y \approx i {\cal B}_0 \> r^n \> e^{i n \varphi}    \nonumber \\
{\rm III)}\> B_x - i B_y &\approx& {\cal B}_0 \> r^n \> e^{-i n \varphi}  \>\>\>\>\>\>\>\>
{\rm IV)} \> B_x - i B_y \approx - i {\cal B}_0 \> r^n \> e^{-i n \varphi} \>.\label{campos}
\end{eqnarray}
The first two obey Maxwell equations, the last two do not; as a matter of fact, case (I) with $n=1$ is commonly produced in a Ioffe-Pritchard trap, and $n=2$ is a quadrupole trap \cite{Leanhardt}.  Figure (\ref{fieldB}) shows the structure of the magnetic fields for the case $n = 1$. Case (III) is the {\it hedgehog} configuration usually considered in Skyrmion discussions \cite{Wilczek-Zee,Shnir}. 
As we will see, the local structure of the transverse magnetic field at its vanishing points plays a fundamental role in the final signs of the topological charges of the vortices and Skyrmions. As discussed in Section V below, such a structure can be associated to a topological quantity, namely, the {\it Index} of the field at the vanishing point. \cite{doCarmo}\\

\begin{figure}[htbp]
\begin{center}
\includegraphics[scale=0.5]{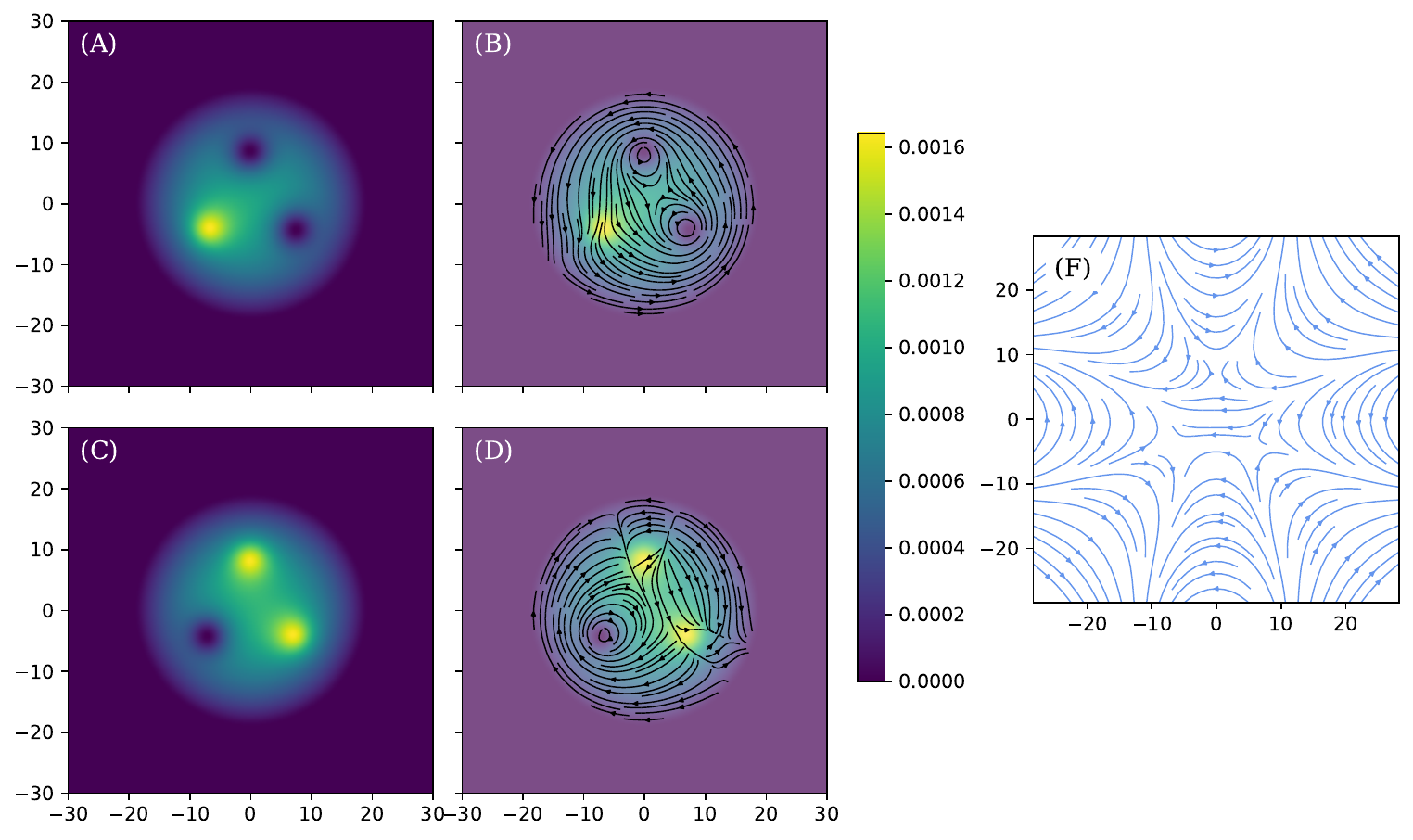}
\end{center}
\caption{Superfluid densities and velocities for the case of figure (\ref{Skyrm-2-1-A}), see Appendix for calculation details. Panels (A) and (B) show the density profile $\rho_+ = |\psi_+|^2$ and the velocity field $\vec v_+$, respectively, with two positive circulation vortices of charge $Q_V = +1$, each one, and a density spike. Panels (C) and (D) show the density $\rho_- = |\psi_-|^2$ and its velocity $\vec v_-$, with a single vortex of negative circulation and charge $Q_V = -1$, and two density spikes. Panel (D) shows the transverse magnetic field lines that gives rise to this configuration. 
 In the vicinity of the vortices the magnetic field is of the type (II) for the upper vortex and type (I) for the lower ones, see (\ref{campos}). See the Appendix for calculational details.}
\label{skyrm}
\end{figure}

Let us consider case (I) for the moment. We will summarize all cases below.  For this case there are two possible solutions,
\begin{eqnarray}
{\rm i)}\> \psi_+(\vec r) &=& F_+(r) e^{-in\varphi} \>\>\>\>{\rm and}\>\>\>\> \psi_-(\vec r) = F_-(r)  \nonumber \\
{\rm ii)} \psi_+(\vec r) &=& F_ -(r)  \>\>\>\>\>\>\>\>\>\>\>\>\>\>{\rm and}\>\>\>\> \psi_-(\vec r) = F_+(r) e^{in\varphi} \>.\label{sols}
\end{eqnarray}
Substitution of these solution into GP equations (\ref{GP2}) shows that, very near $r = 0$, one obtains
\begin{eqnarray}
F_+(r) &\approx& A \> r^n \nonumber \\
F_-(r) &\approx& \frac{\mu}{g} \>,
\end{eqnarray}
where $A$ is a positive constant that can be numerically determined. The important features are that $F_+(r)$ vanishes as $r^n$ while  $F_-(r)$ yields a constant as $r \to 0$. Therefore, case (i) in (\ref{sols}) is a vortex in the $\psi_+$ component and a density ``spike'' in the $\psi_-$ one (see below), and the opposite in (ii), namely a spike in $\psi_+$ and a vortex in $\psi_-$. As we see now, in the former case the vortex charge is $Q_V = - n$, while in the latter $Q_V = + n$. This can be concluded right away from the velocity fields,
\begin{eqnarray}
{\rm i)}\>\> \vec v_+(\vec r) &=& - n \frac{\hbar}{m}\frac{\hat \varphi}{r} \>\>\>\>\>{\rm and}\>\>\>\>\> \vec v_-(\vec r) = 0  \nonumber \\
{\rm ii)}\> \vec v_+(\vec r) &=& 0  \>\>\>\>\>\>\>\>\>\>\>\>\>\>\>\>\>\>{\rm and}\>\>\>\> \vec v_-(\vec r) = + n \frac{\hbar}{m}\frac{\hat \varphi}{r}\>,\label{sols}
\end{eqnarray}
where $\hat \varphi = - \hat x \sin \varphi + \hat y \cos \varphi$. Indeed, the velocity fields circulate in the clockwise (negative) direction in the $\psi_+$ component for case (i), and in the anticlockwise (positive) direction for the $\psi_-$ one, in case (ii). The circulations are, using a circular contour of radius $r$ and $d\vec r = \hat \varphi\>  r d\varphi$,
\begin{eqnarray}
C_{\pm} &=& \oint_{\cal C} \vec v_\pm \cdot d\vec r \nonumber \\
& = & \mp  n \frac{ \hbar}{m} \oint d\varphi \nonumber \\
& = & \mp 2\pi n\frac{\hbar}{m} \>,\label{circ2}
\end{eqnarray}
finding the values of the topological charges already quoted. Clearly, if the vortex is not enclosed by the contour, the circulation vanishes, $C_\pm = 0$. Figure \ref{skyrm} shows the solution to the GP equations (\ref{GP2}), as well as the velocity fields, corresponding to figures (\ref{Skyrm-2-1-A}) and (\ref{Skyrm-2-1-B}), including the magnetic field that generates such a configuration; see the Appendix for details of the calculation. One can observe that, in the vicinity of the vortices, the top one is of type (II) while the lower ones are of type (I) in (\ref{campos}). In panels (A) and (C) of \ref{skyrm} we can also see what we mean by a ``density spike'': while the vortex in one component occurs, accompanied by a density depletion, in the other component there is a rapid increase of the density at the same spatial location, namely a spike, identified in the figure by a bright spot in the density profile. This increment of the density in one spinor component almost cancels the depletion at the vortex in the other component \cite{gallery}.  
From expressions (\ref{campos}) of the fields, the vortex structure of case (II) is the same as case (I) just discussed, however in cases (III) and (IV) the solution is as in (I) given by (\ref{sols}), but with the velocity circulation reversed and, therefore, the charges are now $Q_V = +n$ for solution (i) and $Q_V = -n$ for (ii). As we will see below, the Skyrmion charges will have exactly the same values as $Q_V$, with their own spin texture structure.\\

We now turn our attention to the spin texture and, for the sake of argument, we also study case (I). Again, in the vicinity of zero field, one finds the spin texture from (\ref{f}), for both solutions (i) and (ii) in (\ref{sols})
\begin{eqnarray}
f_x &=& \frac{F_+ F_-}{F_+^2 + F_-^2} \cos n \varphi \nonumber \\
f_y &=& -  \frac{F_+ F_-}{F_+^2 + F_-^2} \sin n \varphi \nonumber \\
f_z & = &  \pm \frac{F_+^2 - F_-^2}{F_+^2 + F_-^2} \>.\label{spinIi}
\end{eqnarray}
where the $\pm$ signs in front of $f_z$ refer to solutions (i) and (ii), respectively, see (\ref{sols}). First, we note that the {\it transverse} spin texture $\vec f_T = f_x \hat x + f_y \hat y$  is parallel and in the same direction to the transverse magnetic field $\vec B_T$, as physically expected, see (I) in (\ref{campos}), thus showing that both have the same spatial pattern. This is true in general and it is exemplified in the gallery of figures in the Appendix.
Evidently, as $r \to 0$, the components $f_x$ and $f_y$ vanish since either $F_+$ or $F_-$ vanish as well, and the spin texture becomes parallel to the $\hat z$ vector, pointing upwards or downwards, $\vec f = \pm \hat z$. This latter direction is determined by whether the solution is (i) or (ii) as above in (\ref{sols}): for solution (i) $|\psi_+| = F_+(0) = 0$, $|\psi_-| = F_-(0) \ne 0$ and, therefore, the spin texture points {\it downwards}, $\vec f(0) = - \hat z$; for solution (ii) $|\psi_-| = F_+(0) = 0$, $|\psi_+| = F_-(0) \ne 0$ and the spin texture now points {\it upwards}, $\vec f(0) = + \hat z$. For the other fields (II), (III) and (IV) the same rule applies for the peaks: if the vortex appears in the $\psi_+$ component, the texture points downwards, but if the vortex is in $\psi_-$ then it points upwards. The ensuing structure of the spin texture is what we call {\it the} Skyrmion and, for the sake of describing it, we shall call the location $\vec r_i$ where $\vec f(\vec r_i) = \pm \hat z$, as the {\it origin} of the Skyrmion. \\

\section{The topological charge of the Skyrmions}

As mentioned in the Introduction, the topological charge of the Skyrmions is usually given by (\ref{Q}) but, as also argued, such an expression requires particular boundary conditions on the spin texture $\vec f$. As also already mentioned, in general those boundary conditions need not occur, or a spin texture can have a multiplicity of coexisting local Skyrmions, whose boundaries among them can also be quite arbitrary. Moreover, in all cases one should be able to asses the individual charges of the local Skyrmions in order to validate that the charges add as any topological index. The spin textures here presented are clear examples of the just two described situations. There is another fundamental aspect to be considered as well. In descriptions of common spin textures and their charge given by (\ref{Q}), it is affirmed that the topological charge is the number of times that the unit sphere is covered by the spin texture. This follows, on the one hand,  by assuming that $\vec f(x,y)$, being a unit vector, is by itself a parametrization or mapping of the unit sphere and, on the other, again, by the mentioned usual boundary conditions. While such a parametrization can indeed be assumed, there is subtle aspect that we believe has not been addressed before, namely, that different spin textures give rise to mappings to the unit sphere but with different orientations. That is, the normal vector to the sphere, generated by the parametrization $\vec f(x,y)$, can point outwards or inwards to the sphere, depending on the given spin texture. And, for configurations with several Skyrmions, this can occur within the same spin texture. This is a very interesting and important point to consider in the elucidation of a general expression for the identification of the Skyrmion topological charge. In this section we present a derivation of the Skyrmion topological charge, in an as possible straightforward fashion, by a procedure similar to the derivation of the Gauss-Bonet (GB) theorem \cite{Weatherburn,doCarmo,Kreyszig}. We will  leave the discussion of the mentioned delicate issues to the following section, where we also argue about the similarities or, rather, dissimilarities of the Skyrmion charge with the Euler-Poincar\'e characteristic $\chi$, arising from the GB theorem.\\

\begin{figure}[htbp]
\begin{center}
\includegraphics[scale=0.25]{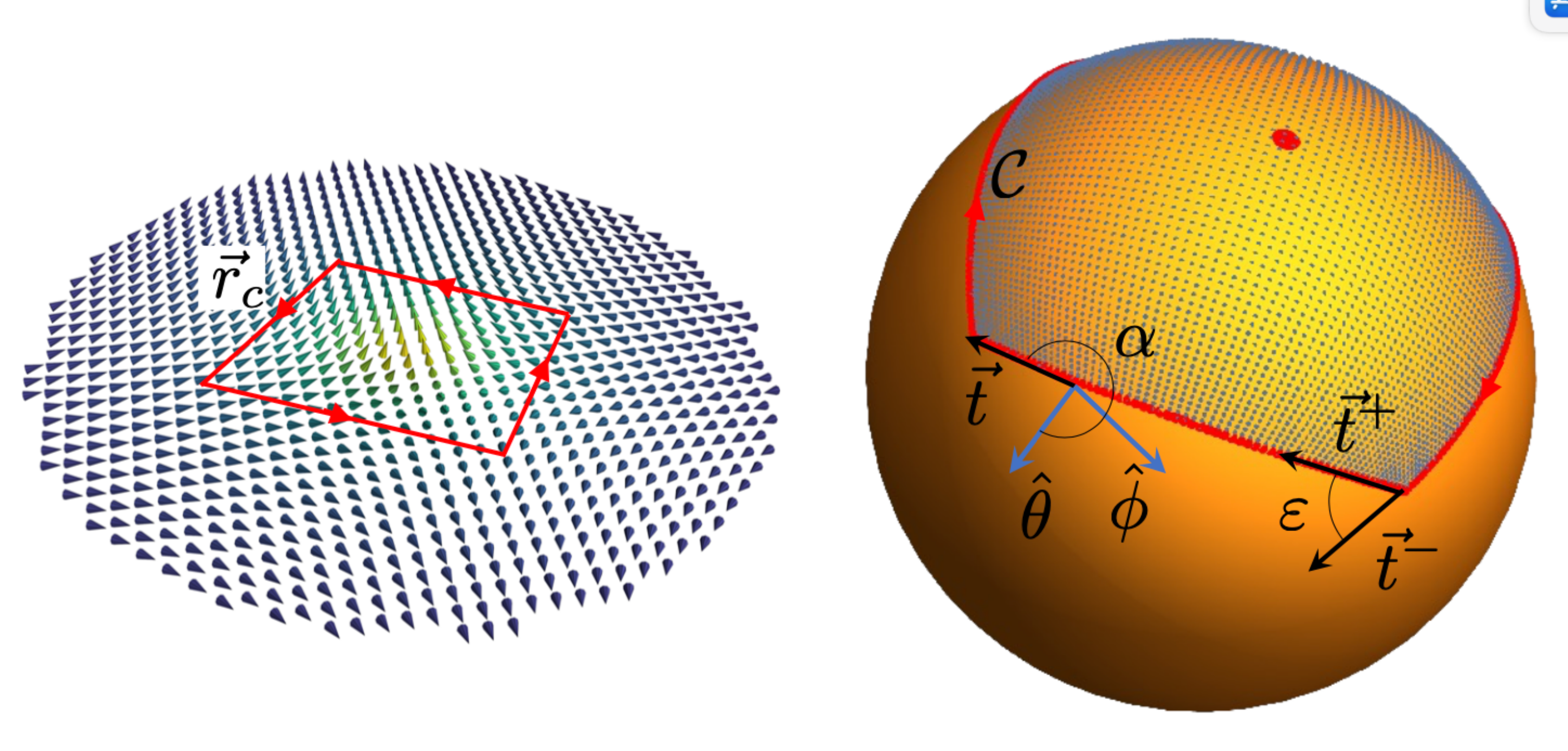}
\end{center}
\caption{Spin texture generated by the magnetic field $\vec B = {\cal B}_0  \left( x \>\hat x - y\> \hat y \right)$ of type (I) in (\ref{campos}), with a Skyrmion origin $\vec f(0) = +\hat z$. On the spin texture, left panel, we show a boundary square curve $\vec r_c$, oriented in the anticlockwise direction. On the right we show the mapping of the spin texture to the unit sphere, within and on the square. The curve on the sphere is ${\cal C}$ found by $\vec f(\vec r_c)$, and results oriented clockwise. The figure also shown a tangent vector $\vec t$ and its angle $\alpha$ relative to the unit sphere vectors $\hat \theta$. The tangent vectors $\vec t^+$ and $\vec t_-$ at one of the corners of the square are also shown, as well as the angle $\varepsilon$ between them. See Appendix for numerical details.
} \label{square}
\end{figure}

Let us assume that a spin texture $\vec f(x,y)$ is given in all or part of the $XY$-plane, such that it has a single location $\vec r_0$ where either $\vec f(\vec r_0) = +z$ or $\vec f(\vec r_0) = -z$. We call this point the {\it origin} of the Skyrmion. Consider now an area $A$ that encloses this point. The area 
has a boundary $\partial A$ described by a parametrized closed curve
\begin{equation}
\vec r_c(t) = \hat x \> x_c(t) + \hat y \> y_c(t) \>\>\>\>\>t : 0 \to T \>\>\>\>\> \vec r_c(0) = \vec r_c(T) \>.\label{ccurve}
\end{equation}
 Notice that although the curve is continuous, its tangent may be discontinuous at a finite number of points $\vec r_c(t_1), \vec r_c(t_2), \dots, \vec r_c(t_s)$. Further, and very important,  the curve is assumed, one, to run or be oriented in the positive counterclockwise direction and, two,  the curve does not cross itself, say the red curve in figure (\ref{square}).\\ 

Evidently, since the texture $\vec f(\vec r)$ is a unit vector, it can be considered to be the outward normal vector $\hat r(\theta,\phi)$  to the unit sphere $S^2$, with $(\theta,\phi)$ the usual azimuthal and polar sphere angles, (please note that we use $\phi$ for the polar angle in 3D and $\varphi$ in 2D)
\begin{eqnarray}
\cos \theta &=& f_z \nonumber \\
\tan \phi & = & \frac{f_y}{f_x} \label{angles} \>.
\end{eqnarray}
At the same time one could assume that $\vec f(\vec r)$ is by itself a {\it parametrization} of the unit sphere. As discussed in detail below why, we will not consider this point of view. As we show below, perhaps surprisingly at  first sight, the two mentioned assumptions are not necessarily consistent with each other. Nevertheless, any value of the spin texture $\vec f(\vec r)$ can be identified with a single point $(\theta,\phi)$ on the sphere, although the opposite is not true in general.  That is, depending on the order of the Skyrmion, parts of the sphere can be covered more than once. Figure (\ref{square}) shows a simple example of these observations, where we can see how the spin texture within and on the square are mapped to unique points on the sphere. And, with the identification of the unit normal and the spherical angles, the usual orthogonal sphere unit vectors $(\hat r, \hat \theta, \hat \phi)$ along the directions $(r,\theta,\phi)$, can also be expressed in terms of the spin texture,
\begin{eqnarray}
\hat r & = & \hat x \> f_x + \hat y \> f_y + \hat z \> f_z \nonumber \\
\hat \theta & = & \hat x \> \frac{f_z f_x}{\sqrt{f_x^2+f_y^2}}+ \hat y \> \frac{f_z f_y}{\sqrt{f_x^2+f_y^2}} - \hat z \> \sqrt{f_x^2+f_y^2} \nonumber \\
\hat \phi & = & -\hat x \>\frac{f_y}{\sqrt{f_x^2+f_y^2}}+ \hat y \> \frac{f_x}{\sqrt{f_x^2+f_y^2}}\>.\label{rthetaphi}
\end{eqnarray}

In an analogous way, we can observe that the curve $\vec r_c(t)$ on the plane gets mapped to a 3D curve $\vec f(\vec r_c(t))$, denoted as ${\cal C}$, that lies on the surface of the unit sphere. The arclength of such a curve can be found from
\begin{equation}
ds = \left|\frac{d\vec f(\vec r_c(t))}{dt}\right| dt \label{ds} \>,
\end{equation}
and the tangent $\vec t(s)$ to the curve is given by, see again figure (\ref{square}) as an example, 
\begin{equation}
\vec t(s) = \frac{d\vec f(s)}{ds} \>.\label{tang} 
\end{equation}
This is a unit vector, $\vec t \cdot \vec t = 1$. Since it is also tangent to the sphere, then $\vec t \cdot \hat r = 0$ and, therefore, $\vec t$ lies on the $\hat \theta$-$\hat \phi$ plane. We define  $\alpha$ as the positive angle that $\hat t$ makes with $\hat \theta$ and, thus, we can write
\begin{equation}
\vec t = \hat \theta \cos \alpha + \hat \phi \sin \alpha \>. \label{alfa}
\end{equation}
$\alpha$ can take any positive value, see figure \ref{square}. Note again that for every texture vector $\vec f$ on the curve ${\cal C}$ there is an angle $\alpha = \alpha(\vec f)$, but the inverse is not necessarily true.\\

With the above preliminaries we can now start the derivation of the expression for the topological charge $Q_S$ of a single Skyrmion. Consider the following integral along the curve ${\cal C}$,
\begin{equation}
I = \oint_{\cal C} \vec f(s) \cdot \left(\vec t(s) \times \frac{d\vec t(s)}{ds}\right) ds\>.\label{I}
\end{equation}
Before we proceed with the evaluation of the integral, we note two points. First, while the integrand may seem to be the {\it geodesic} curvature of ${\cal C}$ on the sphere, we will see that it is not in general, differing by a minus sign when it is not. And second, although the curve $\vec r_c(t)$ on the plane has been considered to run in the positive counterclockwise direction in the $XY$-plane, the curve ${\cal C}$ on the sphere, in general, is not necessarily oriented in the same direction. These apparent discrepancies, to be discussed in the Section V, have their origin in the previous observation that $\vec f$ is not being formally considered to be a parametrization of the sphere.\\

 Substituting the tangent vector (\ref{alfa}) into the integral (\ref{I}) yields,
\begin{equation}
\oint_{\cal C} \vec f(s) \cdot \left(\vec t(s) \times \frac{d\vec t(s)}{ds}\right) ds = \oint_{\cal C} d\alpha + \oint_{\cal C} \cos \theta \> \nabla \phi \cdot d\vec s \>.\label{I2}
\end{equation}
Let us analyze the first integral over the $\alpha$ angle. Here it is important to notice that, because of the assumptions on the boundary curve $\vec r_c(t)$, the tangent $\vec t$ to the curve ${\cal C}$ is also discontinuous at the points $(\vec f(\vec r_1), \vec f(\vec r_2), \cdots, \vec f(\vec r_s))$. This, in turn, implies that the angle $\alpha$ is discontinuous at those  points: Let $\vec t_i^{\> -}$ be the {\it incoming} tangent of the curve at the point $\vec f(\vec r_i)$ and  $\vec t_i^{\> +}$ the {\it outgoing} one. We call the corresponding angles $\alpha_i^-$ and  $\alpha_i^+$. Hence, 
\begin{eqnarray}
\oint_{\cal C} d\alpha &=& \int_{\alpha(0)}^{\alpha_1^-} d\alpha + \int_{\alpha_1^+}^{\alpha_2^-} d\alpha + \cdots + \int_{\alpha_s^+}^{\alpha(T)} d\alpha \nonumber \\
& = & \alpha(T) - \alpha(0) - \sum_{i=1}^s \varepsilon_i\>,\label{dalfa}
\end{eqnarray}
where $\alpha(0) = \alpha(\vec f(0))$ and $\alpha(T) = \alpha(\vec f(T))$ with $\vec f(0) = \vec f(T)$. We have assumed that such a point is not on any discontinuity. $\varepsilon_i = (\alpha_i^+ - \alpha_i^-)$ is the angle between the tangents $\vec t_i^{\> +}$ and $\vec t_i^{\> -}$ at the discontinuity, given by
\begin{equation}
\sin \varepsilon_i = \vec f_i \cdot \left(\vec t_i^{\> -} \times \vec t_i^{\> +}\right) \>,\label{eps}
\end{equation}
see figure \ref{square}. 
Below we return to the analysis of the values of $\alpha(T) - \alpha(0)$. \\

The second integral in (\ref{I2}), over the angle $\phi$, can be rewritten in a way that allows for the determination of another contribution of the Skyrmion charge, using Green theorem. For this, we first recall that the origin $\vec r_0$ of the Skyrmion occurs where the spin texture becomes $\vec f(\vec r_0) = \pm \hat z$, that is, where the azimuthal angle equals $\theta = 0$ or $\pi$, respectively. Since the contour ${\cal C}$ encloses either of those points, the integrand of the second integral in (\ref{I2}) is not analytic at $\theta = 0$ or at $\pi$, preventing the direct use of Green theorem. However, adding and subtracting an appropriate quantity one obtains 
\begin{equation}
\oint_{\cal C} \cos \theta \> \nabla \phi \cdot d\vec s = - \oint_{\cal C} \left(\pm 1 - \cos \theta\right) \> \nabla \phi \cdot d\vec s \pm  \oint_{\cal C} \> \nabla \phi \cdot d\vec s \label{I6} \>,
\end{equation}
where the $+$ sign is used when the singularity is at $\theta = 0$, namely $\vec f(\vec r_0) = + \hat z$, and the $-$ sign when $\theta = \pi$ or $\vec f(\vec r_0) = - \hat z$. In this way, the integrand of the first integral on the right side of (\ref{I6}) is analytic at the respective points and the integral can be performed using Green Theorem. This yields, {\it up to a sign}, the solid angle subtended by the contour ${\cal C}$ around the North or the South Pole,
\begin{equation}
\oint_{\cal C} \left(\pm 1 - \cos \theta\right) \> \nabla \phi \cdot d\vec s  = \pm \iint_{\cal C} d\Omega \>.
\end{equation}
The sign on the right side depends on the positive or negative orientation of the contour ${\cal C}$, which in turn depends on the structure of the Skyrmion in the vicinity of the singularity. The above term can also be seen as a Berry phase type contribution \cite{Berry}. However, the previous integral can be written in a manner that is more appropriate for our purposes, namely, by rewriting it in terms of the spin texture itself:
\begin{eqnarray}
\oint_{\cal C} \left(\pm 1 - \cos \theta\right) \> \nabla \phi \cdot d\vec s 
& = &  \oint_{\vec r_c} \left(\pm 1 - f_z \right)  \nabla \left(\arctan \frac{f_y}{f_x}\right) \cdot d \vec r \nonumber \\
& = & \iint_{A} \left\{ \frac{\partial}{\partial x} \left[ f_z \frac{\partial \arctan \frac{f_y}{fx}}{\partial y}\right] -
\frac{\partial}{\partial y} \left[ f_z \frac{\partial \arctan \frac{f_y}{fx}}{\partial x}\right] \right\} dxdy \nonumber \\
& = & \iint_A \> \vec f \cdot \left(\frac{\partial \vec f}{\partial x} \times \frac{\partial \vec f}{\partial x} \right) \>dx dy \>.\label{surfint}
\end{eqnarray}
The first equality follows by writing the angles $\theta$ and $\phi$ in terms of the $\vec f$ components, see (\ref{angles}), hence allowing for expressing the line integral along the original contour $\vec r_c(t)$. The second equality is obtained by means of Green Theorem, the surface integral being performed over the whole area $A$ enclosed by the curve $\vec r_c(t)$. The derivation of the last equality is a laborious exercise that requires the use of $\vec f \cdot \vec f = 1$. Note that in the expression of the last line one does not need to worry about the sign of the integral, as this will emerge naturally from the current spin texture.  The important issue of these signs will be further addressed below. \\

Collecting the previous results (\ref{dalfa}), (\ref{I6}) and (\ref{surfint}) into (\ref{I2}), we partially obtain
\begin{eqnarray}
\alpha(T) - \alpha(0) \pm  \oint_{\cal C} \> \nabla \phi \cdot d\vec s  &=&  \iint_A \> \vec f \cdot \left(\frac{\partial \vec f}{\partial x} \times \frac{\partial \vec f}{\partial x} \right) \>dx dy + \oint_{\cal C} \vec f(s) \cdot \left(\vec t(s) \times \frac{d\vec t(s)}{ds}\right)ds \nonumber \\
&& + \sum_{i=1}^N \arcsin \vec f_i \cdot \left(\vec t_i^{\> +} \times \vec t_i^{\> -}\right)  \>.\label{QS-2}
\end{eqnarray}
First of all, it is simple to conclude that the left-hand-side of this expression must be 
\begin{equation}
2\pi Q_S = \alpha(T) - \alpha(0) +  \oint_{\cal C} \> \nabla \phi \cdot d\vec s \label{QS-3}
\end{equation}
 with $Q_S \in \mathbb{Z}$. This integer will be carefully  identified below as the topological charge of the Skyrmion enclosed by the curve $\vec r_c$. Quite generally, it must be an integer since, first, $\vec f(0) = \vec f(T)$, then $\sin \alpha(\vec f(0)) = \sin \alpha(\vec f(T))$ and, hence, $\alpha(T) - \alpha(0) = 2 \pi {\cal N}$ with ${\cal N}$ an integer. Similarly, since
\begin{eqnarray}
\pm \oint_{\cal C} \> \nabla \phi \cdot d\vec s &=&\pm \oint_{\cal C} \> d \phi \nonumber\\
& = & 2 \pi {\cal M}\>,\label{m}
\end{eqnarray}
with ${\cal M}$ a positive or negative integer, since the contour ${\cal C}$ encloses either the North or the South Pole of the unit sphere. Therefore, $Q_S = {\cal N} + {\cal M}$ is an integer. However, it remains to understand how this integer number relates to the given spin texture $\vec f(x,y)$ and to its behavior in the neighborhood of the singularity $\vec f(\vec r_0) = \pm \hat z$. \\

\begin{figure}[htbp]
\begin{center}
\includegraphics[scale=0.4]{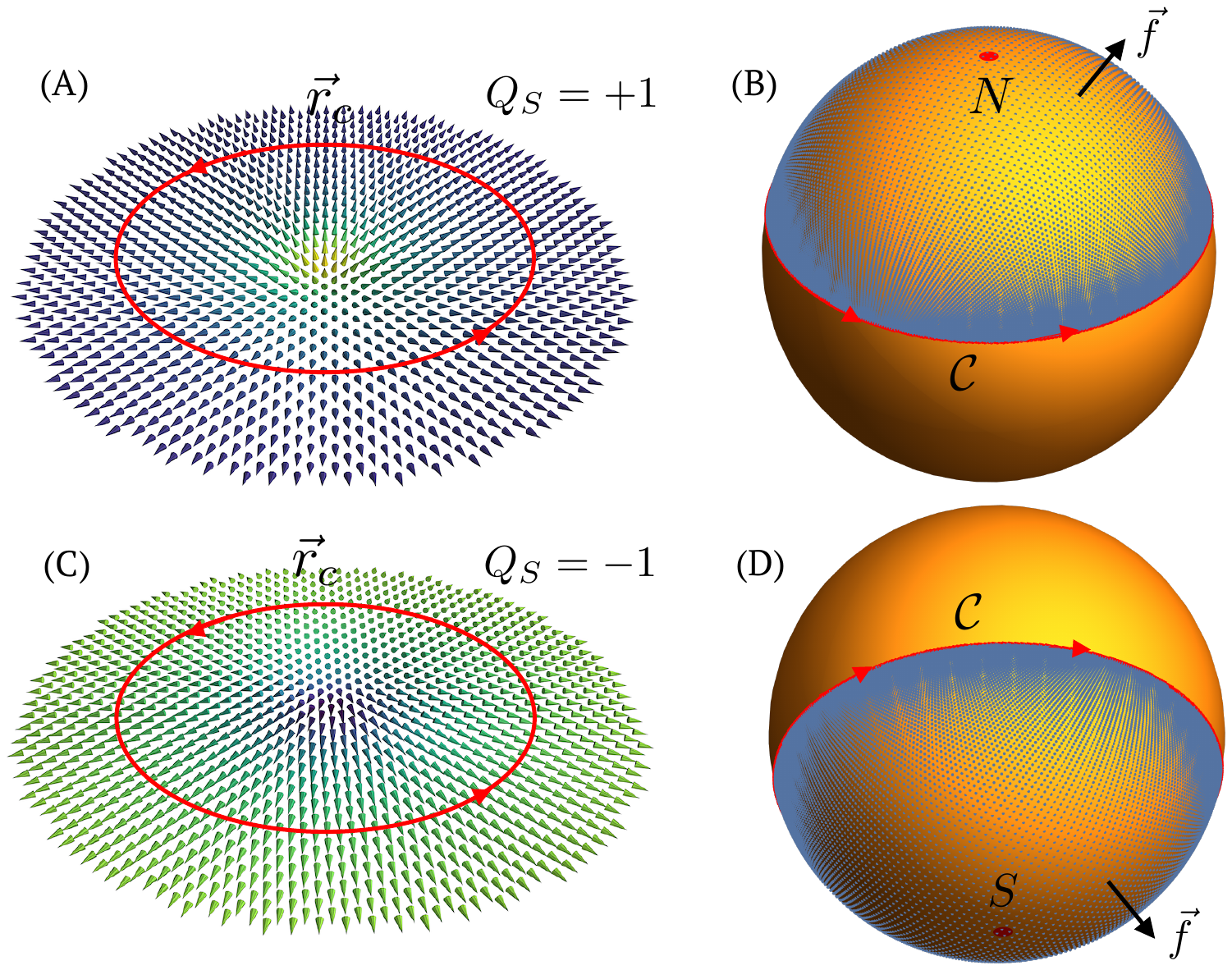}
\end{center}
\caption{Spin texture generated by a  {\it hedgehog} magnetic field $\vec B = {\cal B}_0 \left( x \>\hat x + y\> \hat y \right) + B_z \hat z$, type (III) in (\ref{campos}), and its mapping to the unit sphere. In Panels (A)-(B) $B_z > 0$, the spin texture $\vec f(x,y)$ has a singularity $\vec f(0) = +\hat z$, mapped into the Northern Hemisphere, generating a Skyrmion with charge $Q_S = +1$. In Panels (C)-(D) $B_z < 0$, the singularity is $\vec f(0) = -\hat z$, mapped into the Southern Hemisphere with a Skyrmion charge $Q_S = -1$. The transverse spin texture $\vec f_T = f_x \hat x + f_y \hat y$ is the same in both cases, with an Index $I = +1$. The (red) boundary curve $\vec r_c$ in both spin textures is positively oriented and maps into a curve ${\cal C}$ on the sphere, also positively oriented. An actual numerical calculation of $Q_S$ using (\ref{QS-final}), with a circle of radius $R = 10$, gives $Q_2 \approx \pm 0.712$ and $Q_1 \approx \pm 0.292$, yielding $Q_S = Q_2+Q_1 \approx \pm 1.004$ respectively. See the Appendix for further details of the calculation, here $\mu_0 {\cal B}_0 = 0.1$ and $\mu_0 B_z = \pm 3.0$; the chemical potential in both cases is $\mu = 20.0773$.
} \label{Sky-sph-pm}
\end{figure}

Let us address the role of the angles $\alpha$ and $\phi$ in the determination of the integer value of the Skyrmion charge $Q_S$ in (\ref{QS-3}). First, one can conclude that $\alpha(0) = \alpha(T)$, namely ${\cal N} = 0$. 
This follows from expression (\ref{alfa}) for $\alpha$ and from the geometrical facts, one, that 
$\hat  \theta$ always points in the direction of increasing $\theta$, two, that the tangent vector $\vec t$ has a definite orientation, counter or clockwise for any given curve ${\cal C}$ and, three, that either the North or South Pole are enclosed by ${\cal C}$.
These three facts imply that the angle $\alpha$ retraces itself around ${\cal C}$ without completing a single turn. Thus, this leaves the value of $Q_S$ to depend solely on the integral of $\phi$ in (\ref{QS-3}) and (\ref{m}). This has an intrinsic topological solution. For this, we write the integral over $\phi$ in terms of an integral of the spin texture over the original contour $\vec r_c(t)$,
\begin{equation}
\pm \oint_{{\cal C}_i} d\phi = \pm \int_{\vec r_c^{(i)}} \nabla \left(\arctan \frac{f_y}{f_x}\right) \cdot d \vec r_c^{\>(i)} \>.
\end{equation}
The interesting and relevant observation if that the integral on the right-hand-side is precisely the {\it Index} $I(\vec f_T(\vec r_0))$ of the {\it transverse} spin texture field $\vec f_T(\vec r)$, times $2\pi$, around $\vec r_0$ \cite{doCarmo}. Such an Index is $I = \pm n$ with $n$ a positive integer that counts the number of turns the angle $\phi$ does around the corresponding Pole. 
Based on our findings that the  transverse spin texture $\vec f_T(\vec r)$ follows the transverse magnetic field $\vec B_T(\vec r)$, we can extrapolate and assert that any spin texture, in the vicinity of the singularities $\vec f_T(\vec r_0)$ = 0, behaves just as the magnetic field given in (\ref{campos}). Hence, for cases (I) and (II) the index is $I = -n$ while for (III) and (IV) one obtains $I = +n$. Therefore, in a very general way, the topological charge of a single Skyrmion can be expressed as,
\begin{equation}
Q_S = f_z(\vec  r_0) \> I(\vec f_T(\vec r_0))  \>,\label{Q7}
\end{equation}
demonstrating that the topological charge is a positive or a negative integer, its sign being determined by both, the value of $f_z$ and the Index of $\vec f_T$, at the singularity. We believe that this last identification of the Skyrmion charge has not been pointed out before. Figure \ref{Sky-sph-pm} shows two cases of  Skyrmions with singularities at $\vec f = +\hat z$ and $\vec f = +\hat z$, produced by the same transverse magnetic field but with opposite $B_z$ component; all the thermodynamic properties are the same, except that the Skymion charges have opposite values.\\

\begin{figure}[htbp]
\begin{center}
\includegraphics[scale=0.4]{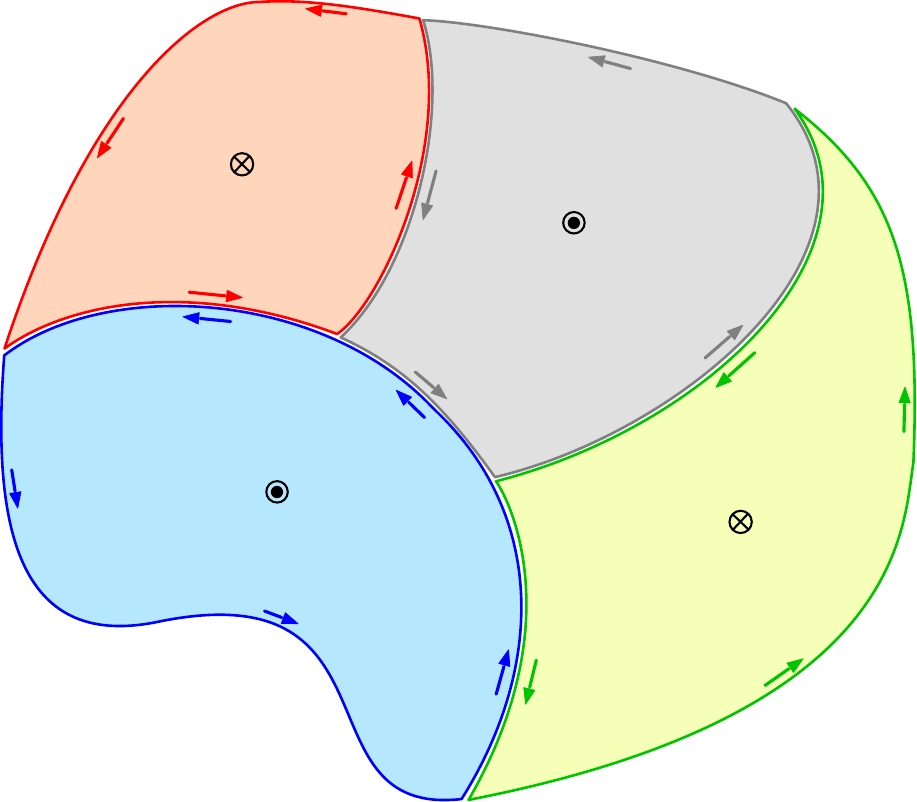}
\end{center}
\caption{Schematic representation of a spin texture configuration with multiple Skyrmions, as the union of local textures holding one Skyrmion each. The black dots represent Skyrmions with $f_z = +\hat z$ and the crosses with $f_z = -\hat z$. Applying expression (\ref{QS-final}) to each of the colored regions yield the local Skyrmion charges and their union equals expression (\ref{QS-final}) to total Skyrmion charge. 
} \label{multiple}
\end{figure}

Returning to the identification of  $Q_S$ in (\ref{QS-2}), and rewriting the integrals in terms of the spin texture $\vec f$ and the original boundary curve $\vec r_c$, one arrives at the sought for expression for the topological Skyrmion charge given in (\ref{QS-final}) in the Introduction, but for a single Skyrmion. The extension to the total topological charge of a spin texture that holds an arbitrary number $N$ of Skyrmions, whose origins are at $\vec r_1, \vec r_2, \dots, \vec r_N$, is straightforward. This is achieved with the 
reasonable assumption that, while in the vicinity of the origins the spin texture behaves as in the case of a single Skyrmion, the ``local'' textures can be  joined smoothly in a continuous and differentiable fashion. This assumption is verified by the numerical examples shown in the Appendix.
Very importantly, the boundaries between those local cases can be quite arbitrary and complicated, such as the boundary of our example of the three Skyrmions in figure (\ref{Skyrm-2-1-A}). This corroborates the importance of having obtained the local charge of a Skyrmion for an arbitrary boundary curve $\vec r_c$. Figure \ref{multiple} shows a 
sketch of the union of local boundary curves $\vec r_c^{(i)}$ of several Skyrmions that yields a total curve $\vec r_c$, inside of which all Skyrmions are enclosed. Writing the different contributions of a local Skyrmion charge as $Q_S^{(i)} = Q_2^{(i)} + Q_1^{(i)} +Q_0^{(i)}$, where
\begin{eqnarray}
Q_2^{(i)}  &=&  \iint_{A_i} \> \vec f \cdot \left(\frac{\partial \vec f}{\partial x} \times \frac{\partial \vec f}{\partial x} \right) \>dx dy \nonumber \\
Q_1^{(i)} & = & \oint_{\vec r_c^{\>(i)}} \vec f \cdot 
\frac{\left( \frac{d \vec f}{dt} \times \frac{d^2\vec f}{dt^2} \right) }{ \left|\frac{d \vec f}{dt}\right|^2 }\
dt  \nonumber \\
Q_0^{(i)} & = & \sum_{j=1}^{s^{(i)}} \sin^{-1} \vec f_j \cdot \frac{\left( \left. \frac{d \vec f}{dt}\right|_j^{\> -} \times \left. \frac{d \vec f}{dt}\right|_j^{\> +}\right)}{\left| \frac{d \vec f}{dt}\right|_j^{\> -}\left| \frac{d \vec f}{dt}\right|_j^{\> +} }\>, \label{QS-partes}
\end{eqnarray}
it is evident, since all boundary curves $\vec r_c^{(i)}$ run with same positive orientation, that 
\begin{equation}
Q_S = \sum_{i=1}^N Q_S^{(i)}
\end{equation}
with an analogous expression for all contributions $Q_2$, $Q_1$ and $Q_0$. The obtained expression for the total Skyrmion charge $Q_S = Q_2 + Q_1 + Q_0$ is given by (\ref{QS-final}) of the Introduction, with $\vec r_c = \bigcup_{i=1}^N \vec r_c^{\>(i)}$ the external boundary curve, $A =  \bigcup_{i=1}^N A_i$, the total area, and the sum over the angles is only on those on the external boundary $\vec r_c$ since all the internal angles cancel out, see figure \ref{multiple}. It is also of interest to conclude that the total Skyrmion charge can be written as,
\begin{equation}
Q_S = \sum_{i=1}^N \> f_z(\vec r_i) \> I(\vec f_T(\vec r_i))  \>.
\end{equation}
As we will argue in the following section, this last expression appears as a generalization of Poincar\'e theorem \cite{doCarmo}. \\

In the Appendix we illustrate the results of this section with explicit numerical calculations of several spin textures obtained by solving GP equations (\ref{GP2}) for our case study of a spinor BEC in the presence of an external field.

\section{Dissimilarities between the Skyrmion charge $Q_S$ and the Euler-Poincar\'e characteristic $\chi$}

The expression (\ref{QS-final}) for the Skyrmion topological charge $Q_S$ has a deceivingly similar appearance to the Euler-Poincar\'e characteristic $\chi$ of the Gauss-Bonet theorem of a given 2D surface embedded in 3D \cite{Weatherburn,Kreyszig,McCleary,doCarmo}. As a matter of fact, the derivation here presented for $Q_S$ is similar, but not equal, to the derivation of the Gauss-Bonet theorem presented by do Carmo \cite{doCarmo}. However, as we have been alluding to in the text, there are notable differences between them: A technical difference being that the spin texture vector $\vec f$ cannot always be both the normal vector and the parametrization to the sphere at the same time. A fundamental difference being that in the Gauss-Bonet theorem the main role is played by the surface itself, while in the Skyrmion case the protagonist is the spin texture $\vec f$. \\

To be precise, given an {\it orientable} surface $S$, the GB theorem states that the Euler-Poincar\'e characteristic $\chi$ of a region $R$ of the surface $S$, with a given closed boundary $C$ and with $M$ possible exterior angles $\theta_i$ is, \cite{doCarmo}
\begin{equation}
2\pi \chi = \iint_R K \> dA + \oint_C k_g \>ds + \sum_{i=1}^M \theta_i \label{GB}
\end{equation}
where $K$ is the gaussian curvature at a point on the surface, $k_g$ the geodesic curvature at a point of the curve $C$, the boundary of $R$. The curve $C$ must be oriented in the positive direction. These are intrinsic properties of the surface and of curves drawn on it. If the region $R$ is a simple region of the surface $S$, then $\chi = +1$ always.  If the surface is compact with no boundaries, then $\chi = 2 - 2g$ where $g$ is the genus or number of handles of the surface. And there are surfaces with $\chi = -1$ such as the pseudo sphere and the so-called {\it pair of pants}. The above expression for $\chi$ is written in an invariant form, independent of any parametrization of the surface. For sure, one can write it in terms of a parametrization, but this one must be consistent with an assumed orientation of the surface and with the positive orientation of the curve $C$. To see this, let $\vec X(\sigma_1,\sigma_2)$ be such a parametrization, then, the normal to the surface is
\begin{equation}
\hat n = \frac{\partial_1 \vec X \times \partial_2 \vec X}{\left|\partial_1 \vec X \times \partial_2 \vec X\right|} \label{nor}\>,
\end{equation}
where $\partial_\alpha$ is the derivative with respect to $\sigma_\alpha$. This normal must have the same orientation on the whole surface. For instance, if the surface is a sphere, then $\hat n$ must point outwards or inwards over the whole sphere. The gaussian curvature can be expressed as,
\begin{equation}
K = \hat n \cdot \left(\partial_1 \hat n \times \partial_2 \hat n\right) \>.\label{K}
\end{equation}
Along the same lines, the tangent to the curve $C$ is
\begin{equation}
\vec t = \frac{\partial \vec X(s)}{\partial s} \>,\label{tG}
\end{equation}
and, hence, the geodesic curvature of $C$ is
\begin{equation}
k_g = \hat n \cdot \left(\vec t \times \frac{d\vec t}{ds}\right) \>.\label{kg}
\end{equation}
The exterior angles $\theta_i$ are, in an obvious notation,
\begin{equation}
\sin \theta_i = \hat n \cdot \left( \vec t_i^{-} \times \vec t_i^{+} \right) \>.\label{thetai}
\end{equation}

Therefore, if we compare $\chi$, from (\ref{GB})-(\ref{thetai}), with $Q_S$ from (\ref{QS-final}), they look quite similar if we identify the parametrization $\vec X$ with $\vec f$, the surface thus being a sphere, and the normal $\hat n$ with $\vec f$. However, these two identifications are incompatible in general. This can be readily seen if we calculate the normal to the sphere from the parametrization $\vec X = \vec f$, 
\begin{equation}
\hat n = \frac{\partial_x \vec f \times \partial_y \vec f}{\left|\partial_x \vec f \times \partial_y \vec f\right|} \>.
\end{equation}
One finds that, depending on the spin texture, $\hat n = \vec f$ or $\hat n = -\vec f$. And this can occur within the same spin texture, such as in the example of figure (\ref{Skyrm-2-1-A}). Therefore, if $\vec f$ is a parametrization, then $\vec f$ cannot be the normal always, and viceversa. This would result in severe problems since the normal would discontinuously change on the sphere. Hence, we have opted to consider 
that $\vec f$ is the outward normal to the sphere, at the expense of abandoning $\vec f$  as a parametrization. It is then not a surprise that we do not obtain GB theorem and, therefore, $Q_S$ is not $\chi$, since the sign of the former depends on the spin texture $\vec f$ and the latter is a topological intrinsic property of the given surface. Indeed, the fact that we can find positive or negative Skyrmion charges, simply by inverting the value of the origin of the Skyrmion $\vec f(\vec r_0)$ is a consequence of not having considered $\vec f$ as a parametrization of the sphere. In this way, the integrands of $Q_2$ and $Q_1$ are not necessarily the gaussian or the geodesic curvatures of the sphere or of the curves drawn on it.
It is also of interest to point out that there is a theorem, due to Poincar\'e \cite{doCarmo}, that states that $\chi = \sum_i I_i$ with $I_i$ the Indices of a differentiable vector field on a compact surface with isolated singular points. That is, given the surface, the singularities must exist and their Indices add to $\chi$. In the present case, see (\ref{Q7}), the singularities can be quite arbitrarily given by the the spin texture and, as a consequence, their topological charges depend on an additional sign given by $f_z(\vec r_0)$ at the singularity. \\

The previous observations are not definite, however, as we are aware that topology can yield results that may be difficult to visualize geometrically. That is, although we have pointed out evident differences between the Skyrmion topological charge and the Euler-Poincar\'e characteristic, we also concede that their respective expressions appear strikingly similar. Hence, we leave open the question of whether they may truly be the same topological object or not, perhaps for not so obvious or evident surfaces such as the unit sphere. This should be the subject of further scrutiny. Along the same lines, an approach in terms of homotopy theory \cite{Ueda,Schwarz,Mermin} should complement our approach, basically rooted in differential geometry.

\section{Final Remarks}

In this article we have provided a derivation of the Skyrmion topological charge $Q_S$ of an arbitrary spatial region $A \subset \mathbb{R}^2$ of a spin texture $\vec f(\vec r)$, given by (\ref{QS-final}). If the spin texture is continuous and differentiable everywhere, except perhaps at the origins of the Skyrmions, where $\vec f = \pm \hat z$, one can find the local topological charges of the present Skyrmions and, simply by joining the local regions, one can calculate the topological charge of several Skyrmion regions and, eventually, of the whole area where the spin texture is defined. While we can calculate the topological charge as the sum of the three main contributions $Q_S = Q_2 + Q_1 + Q_0$, one can also demonstrate that the local Skyrmion charge is the product $Q_S = f_z(\vec r_0) I(\vec f_T(\vec r_0))$, with $I$ the Index of the transverse spin texture at the singularity $\vec r_0$. The total Skyrmion charge of a region, or of the whole spin texture, is the addition of the local Skyrmion charges. It may be of formal interest to highlight that the 2D contribution $Q_2$, the 1D $Q_1$ and the 0D $Q_0$ are all expressed in terms of a triple scalar product involving the spin texture $\vec f$ times the relevant cross product of the contribution at hand. We also find that, depending on the value of the spin texture at the boundary curve or curves, one recovers the usual expression for the Skyrmion charge, given solely in terms of the contribution $Q_2$. For instance, if  the texture completely covers the sphere, $\vec f = \pm \hat z$, or $f_z = 0$ at the boundary, the last two terms in (\ref{QS-final}) or (\ref{QS-2}) vanish. Notice also that we have defined the Skyrmion charge with a factor $1/2\pi$ in (\ref{QS-final}), instead of $1/4\pi$ as in (\ref{Q}), because of the similarities  with the Gauss-Bonet representation of the Euler-Poincare characteristic. Thus, while usually the Skyrmion charge for a closed sphere is $Q = \pm 1$, here it is $Q_S = \pm 2$, and as the charge of a half Skyrmion is considered $Q = \pm 1/2$, here it is a $Q_S = \pm 1$. As we indicated in the Introduction, the derived expression is quite general and independent of the origin of the spin texture.\\

In order to produce spin texture within a physically sensible system, we have studied a 1/2 spinor BE condensate in the presence of an external magnetic field. This system, in addition to generate interesting and non trivial spin textures, with which we can make a detailed and systematic verification of our main result, presents quantum vortices, defects of topological nature as well. As we have indicated, one can trace the appearance of the vortices and of the Skyrmions to the behavior of the external magnetic field in the vicinity of the locations $\vec r_0$ where the transverse magnetic field vanishes. Indeed, by looking a the emergence of quantum vortices, say, in (\ref{circ2}), one can conclude that the vortex local topological charge is $Q_V = \pm I(\vec B_T(\vec r_0))$, where the sign $\pm$ is if the vortex appears in the $\psi_+$ or $\psi_-$ spinor component, respectively, and $I(\vec B_T(\vec r_0))$ is the Index of the transverse magnetic field in the vicinity of its vanishing location $\vec r_0$. However, since the transverse spin texture $\vec f_T$ follows exactly the same pattern as the transverse magnetic field $\vec B_T$, see Appendix, then the local quantum vortex and Skyrmion topological charges are exactly the same. That is, for the present case study, the external magnetic field determines the topology of all ensuing structures. These results can definitely be tested with the currently achieved experimental capabilities in the study of spinor Bose-Einstein condensates. For other physical spin textures one can perhaps always find the underlying field or interaction responsible for the emergence of the corresponding topological structures. \\

While we have left out several additional observations of this problem, related to its intrinsic differential geometry and topology, we mention one that should be of importance for further analysis. This regards the issue of what we call a ``Skyrmion'' structure in a spin texture. That is, the emphasis is always in the role of the location where one finds the origin of the Skyrmion $f_z(\vec r_0) = \pm \hat z$ and, therefore, one may believe that the determination of the topological charge needs of a boundary curve that encloses such a point or points. However, this is not completely true or necessary. For this we point out expression (\ref{QS-3}) where we identified the Skyrmion charge in terms of the angle difference $\alpha(T)-\alpha(0)$ and the contour integral over $\phi$. We recall that we derived our results by assuming a contour that enclosed the singularity yielding  $\alpha(T)-\alpha(0) = 0$ and the topological charge was contained in the $\phi$ contour integral. However, if we choose a contour that does not enclose the singularity, then the Index of the transverse spin texture vanishes, but now $\alpha(T)-\alpha(0) \ne 0$. And, for sure, the topological charge is now accounted for by the angle $\alpha$, namely $2\pi Q_S = \alpha(T)-\alpha(0)$. This indicates that if we take a fully arbitrary boundary curve $\vec r_c$, yet closed and positively oriented, that does not necessarily enclose singularities, one still finds the topological charge contained within the encircled spin texture. In other words, the topological content of the Skyrmion is within the whole spin texture, as Topology may have indicated us in advance.\\

\medskip\noindent
{\bf Acknowledegment.} This work was partially funded by grant PAPIIT-UNAM IN117623. VRR acknowledges fruitful discussions with Leonardo Pati\~no at the early stages of this work.

\newpage
\section*{Appendix: A gallery of vortices and Skyrmions, solutions to GP equations (\ref{GP2}).}

The figures throughout the text and in this Appendix are found from numerical solutions of the two coupled Gross-Pitaevskii equations (\ref{GP2}). We use a standard method described in Refs. \cite{gallery,chinos}, in a GPU processor, within a $256 \times 256$ grid. We consider arbitrary dimensionless units such that the solutions are numerically stationary and the desired configuration is clearly observed. We use $\hbar = m = 1$, the interatomic coupling $g = 15000$ and, in all cases, the computational box is a square of side $L_0 = 56.72$. The values of the isotropic frequency trap $\omega$, the magnetic fields and the corresponding chemical potential $\mu$ are given in the figures captions. \\

 Recalling that the full magnetic field is $\vec B = \vec B_T + B_z \hat z$, an interesting point to highlight is that both solutions (i) and (ii) in (\ref{sols}) are possible in each position where the transverse magnetic field vanishes. However, if $B_z = 0$, we find numerically that these solutions appear at random, each 50\% of the time, because our numerical methods requires a random guess of the initial wavefunctions \cite{gallery}. However, if we turn on the constant $B_z$ field, we find, as expected, that the sign of this field component forces one of the solutions, namely, if $B_z > 0$, then solution (i) occurs, while (ii) ensues for $B_z < 0$. \\

\begin{figure}[htbp]
\begin{center}
\includegraphics[scale=0.5]{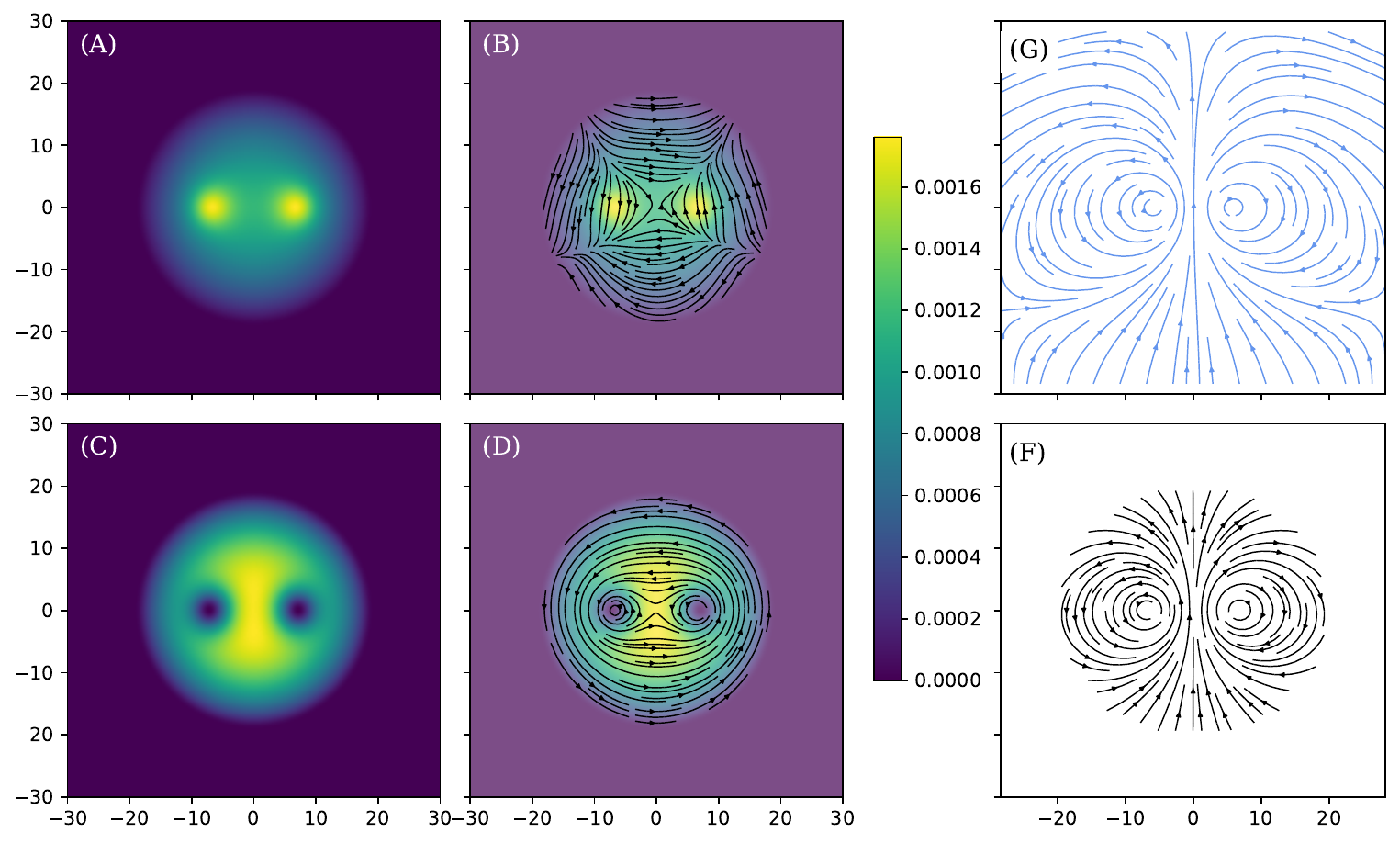}
\includegraphics[scale=0.33]{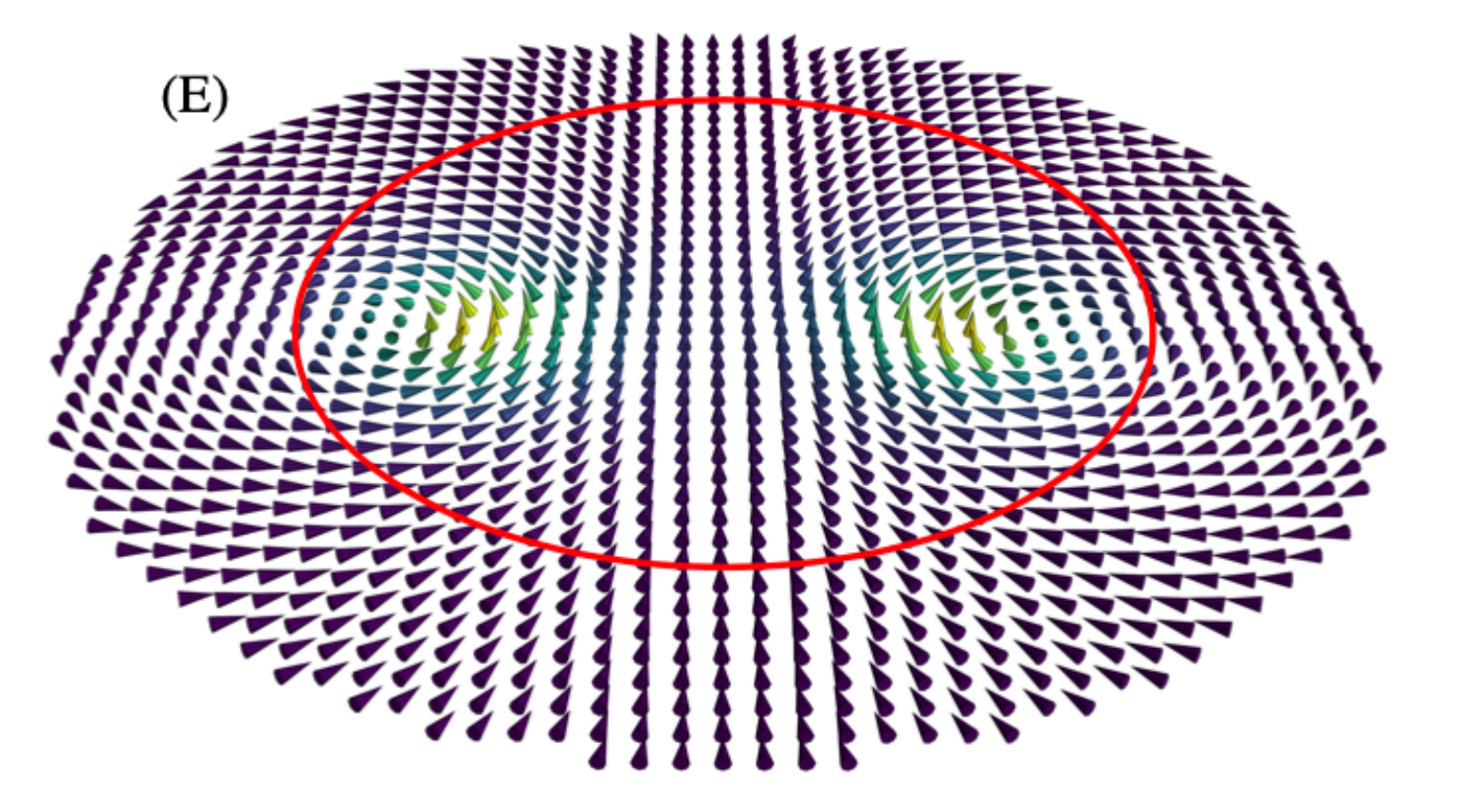}
\end{center}
\caption{Two vortices of charge $Q_V = +1$, panels (A)-(D), in $\psi_-$ component with a positive anticlockwise circulation. Spin texture Syrmion of total charge $Q_S = +2$, the result of adding two local Skyrmions of charge $Q_S = +1$, panels (E)-(F). The Skyrmion charge for the circular boundary curve of radius $R = 10$ in (E) using (\ref{QS-final}) yields $Q_2 \approx 1.978$ and $Q_1 \approx 0.020$, adding to  $Q_S = Q_2 + Q_1 \approx 1.998$. The frequency trap is $\omega^2 = 0.2$ and the chemical potential $\mu = 30.6671$. The magnetic field is obtained by superposing the fields of 3 wires, see (\ref{wire}), with currents $\mu_0 I_1/c = \mu_0 I_2/c = \mu_0 I_3/c = 20.0$ and located at $(x_1,y_1) = (30.0,-16.334)$, $(x_2,y_2) = (0.0,32.666)$ and $(x_3,y_3) = (-30.0,-16.334)$.  Panel (G) shows the transverse field lines $\vec B_T$.
} \label{skyrm-1-1}
\end{figure}

\begin{figure}[htbp]
\begin{center}
\includegraphics[scale=0.5]{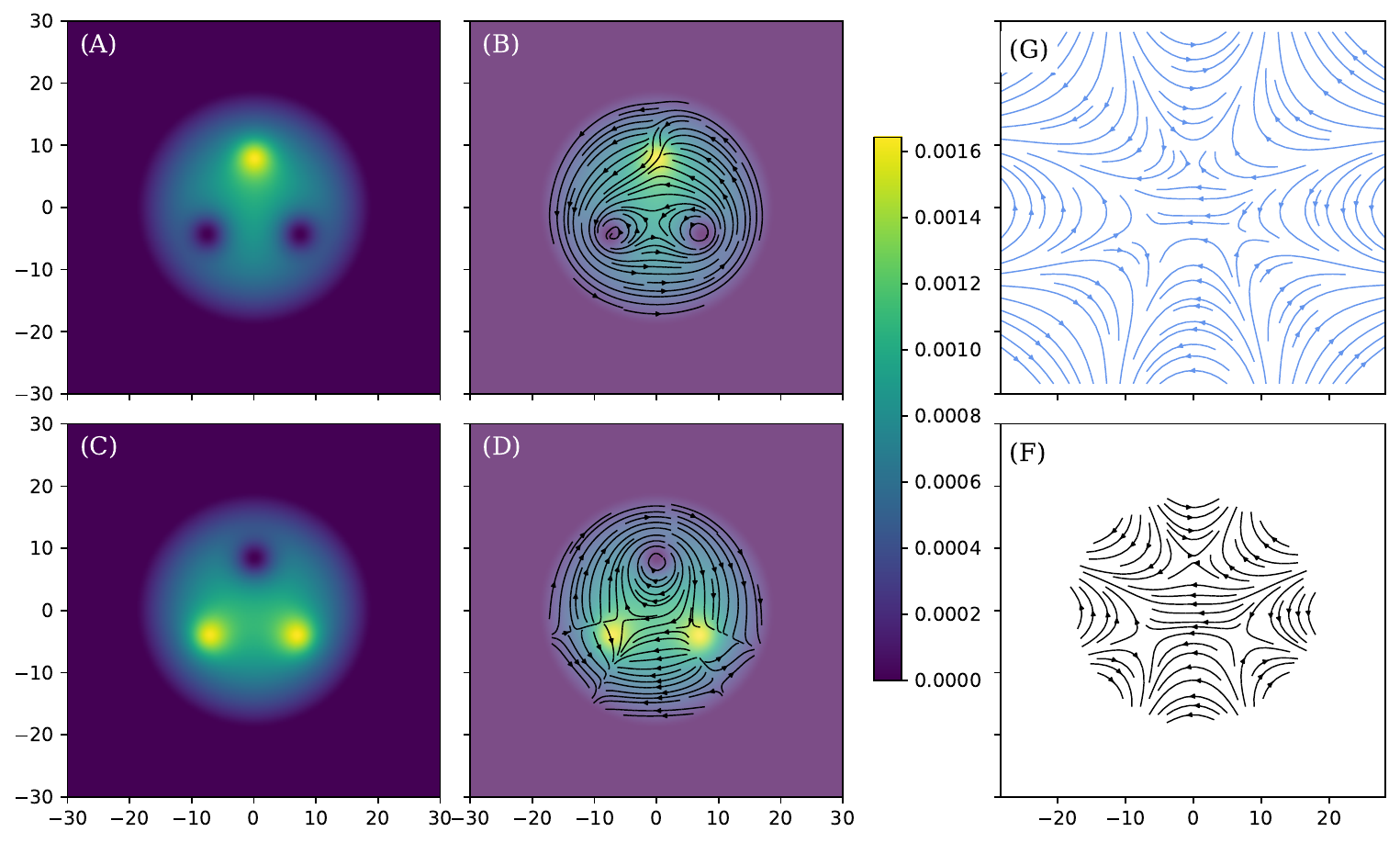}
\includegraphics[scale=0.25]{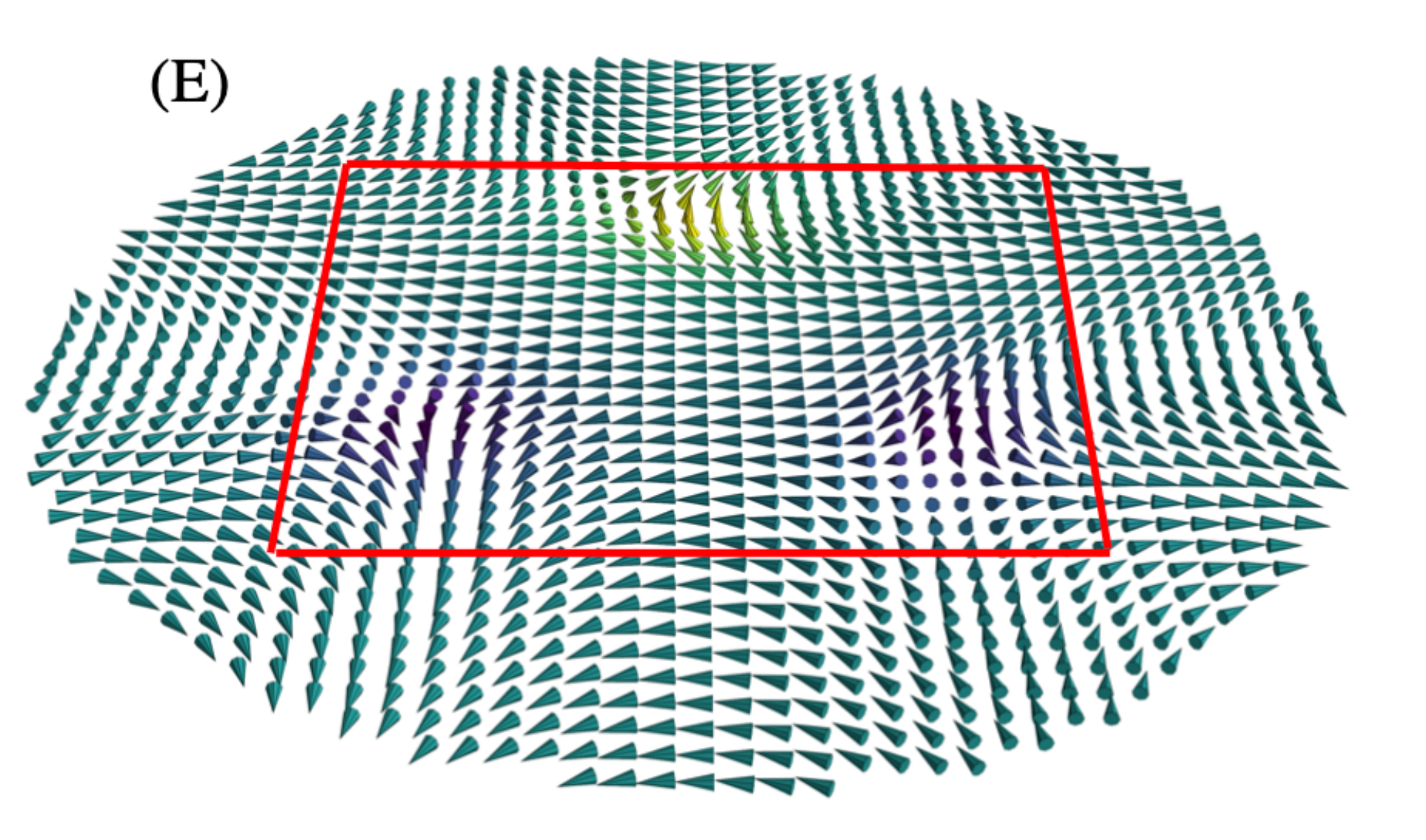}
\end{center}
\caption{
Three vortices, two of charge $Q_V = +1$ in the $\psi_+$ component and one of charge $Q_V = -1$ in the $\psi_-$ one, panels (A)-(D)
Spin texture Syrmion of total charge $Q_S = +1$, the result of adding two of charge $Q_S = +1$ and one of charge $Q_S = -1$, panels (E)-(F). The Skyrmion topological charge for the square boundary in (E) using (\ref{QS-final}) yields $Q_2 \approx 0.934$, $Q_1 \approx 0.023$ and $Q_0 \approx 0.042$ , adding to  $Q_S = Q_2 + Q_1 +Q_0 \approx 0.999$. The frequency trap is $\omega^2 = 0.2$ and the chemical potential $\mu = 30.6041$. The magnetic field is obtained by superposing the fields of 6 wires, see (\ref{wire}), with currents $\mu_0 I_1/c = \mu_0 I_2/c = \mu_0 I_3/c = \mu_0 I_4/c = 76.72, \mu_0 I_5/c = \mu_0 I_6/c = 1.918$ and located at $(x_1,y_1) = (L_0,0)$, $(x_2,y_2) = (0,L_0)$, $(x_3,y_3) = (-L_0,0)$, $(x_4,y_4) = (0,-L_0)$, $(x_5, y_5) = (-L_0,-L_0)$, $(x_5, y_5) = (L_0,-L_0)$, with $L_0 = 38.36$.  Panel (G) shows the transverse field lines $\vec B_T$.} \label{skyrm-2-1}
\end{figure}

The solutions of figures \ref{Skyrm-2-1-A} and \ref{Skyrm-2-1-B}, as well as those below \ref{skyrm-1-1} and \ref{skyrm-2-1}, have been obtained with transverse magnetic fields that become zero at appropriate locations on the $XY$-plane within the BE condensate. These fields are in turn tailored by superposing magnetic fields of a set of long (infinite) wires at appropriate locations, carrying also an appropriate current $I$. We recall that the field of a long straight wire in the $\hat z$-direction, located at  $(x_0,y_0)$, with current $I$, is
\begin{equation}
\vec B_{\rm wire}(\vec r) = \frac{2I}{c}  \> \frac{-(y-y_0) \hat x + (x-x_0) \hat y}{(x-x_0)^2 + (y-y_0)^2} \>.\label{wire}
\end{equation}
In the figures captions we indicate the values of the currents and the locations of the wires. The wires are placed outside of the numerical calculation box.\\

In the following gallery of figures we show the condensate superfluid density $\rho_\alpha$ and velocity $\vec v_\alpha$, for component $\alpha = +$, panels (A) and (B); and for component $\alpha = -$, panels (C) and (D), respectively. In panel (E), we show the Skyrmion spin texture and in panel (F) the transverse spin texture field lines $\vec f_T = f_x \hat x + f_y\hat y$. In panel (G) we show the transverse magnetic field lines $\vec B_T =  B_x \hat x +  B_y \hat x$ that produces the corresponding condensate. Note that the transverse spin texture streamlines follow those of the transverse magnetic field. Details of the corresponding cases are given in the figures captions. \\

\begin{figure}[htbp]
\begin{center}
\includegraphics[scale=0.5]{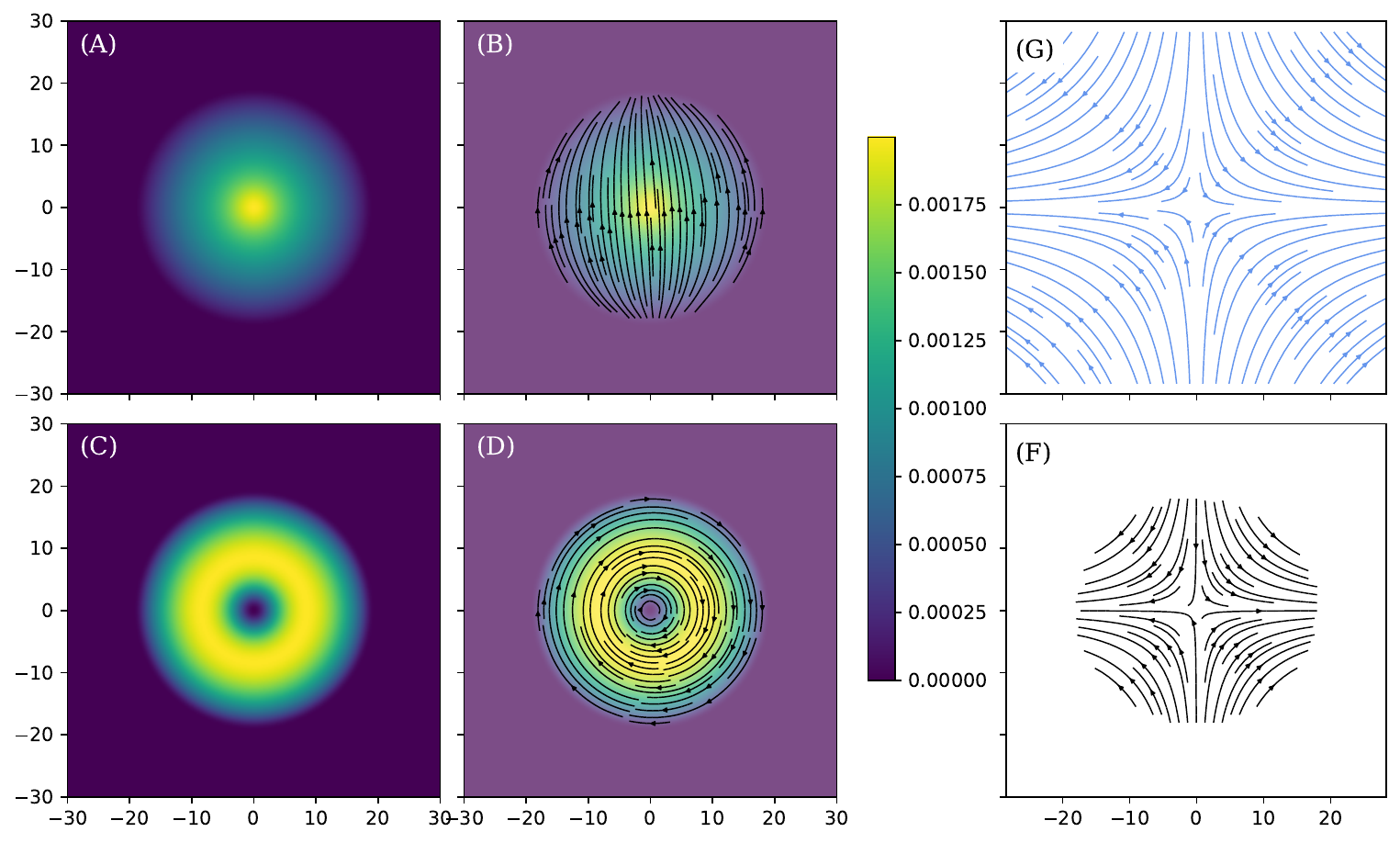}
\includegraphics[scale=0.25]{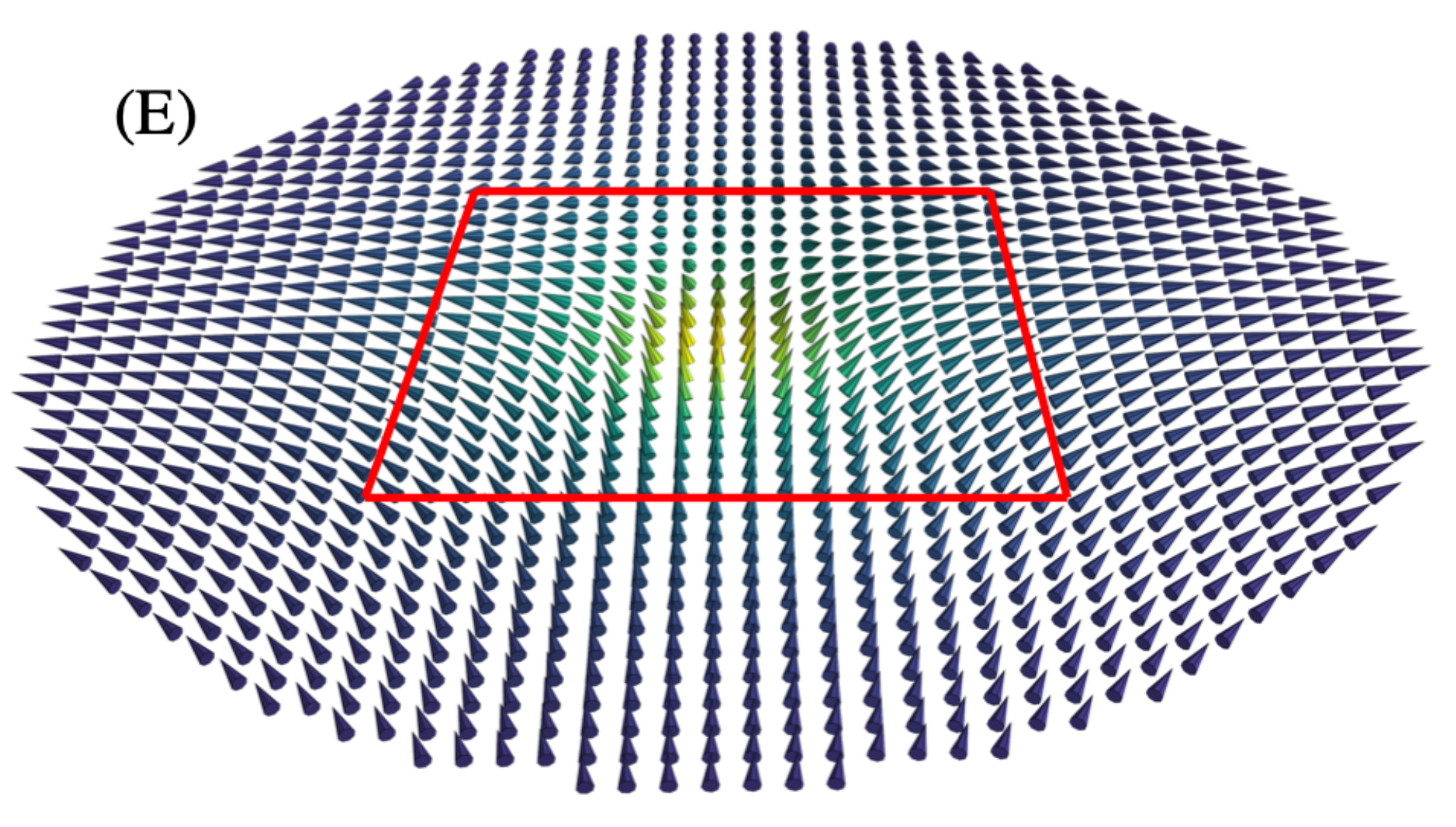}
\end{center}
\caption{A single vortex of charge $Q_V = -1$, panels (A)-(D), in $\psi_-$ component with a negative clockwise circulation. Spin texture Syrmion of charge $Q_S = -1$, panels (E)-(F). Note that although the Skyrmion at its origin points in the $+\hat z$ direction, its topological charge is negative.The Skyrmion charge for the square boundary curve in (E) using (\ref{QS-final}) yields $Q_2 \approx -0.749$, $Q_1 \approx -0.002$ and $Q_0 \approx -0.248$, adding to $Q_S = Q_2 + Q_1 + Q_0 \approx -0.999$. The frequency trap is $\omega^2 = 0.2$ and the chemical potential $\mu = 29.9046$. The magnetic field is $\vec B_T = \mu_0 {\cal B}_0 (x \hat x - y \hat y)$, panel (G), and $B_z = \mu_0 B_0$, with $\mu_0 {\cal B}_0 = 0.1$ and $\mu_0 B_0 = 0.3$.
} \label{skyrm-1}
\end{figure}

\newpage
\begin{figure}[htbp]
\begin{center}
\includegraphics[scale=0.5]{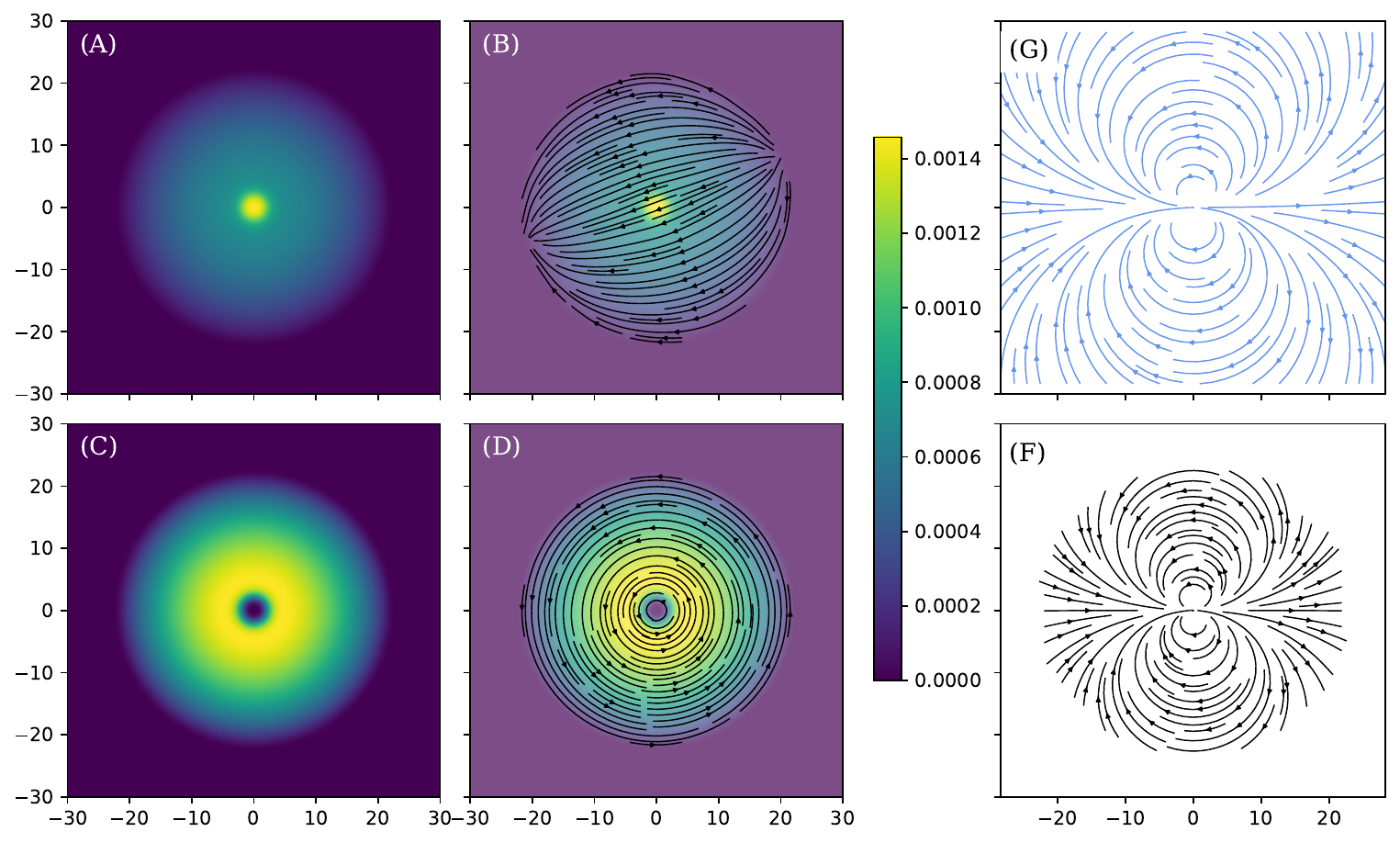}
\includegraphics[scale=0.25]{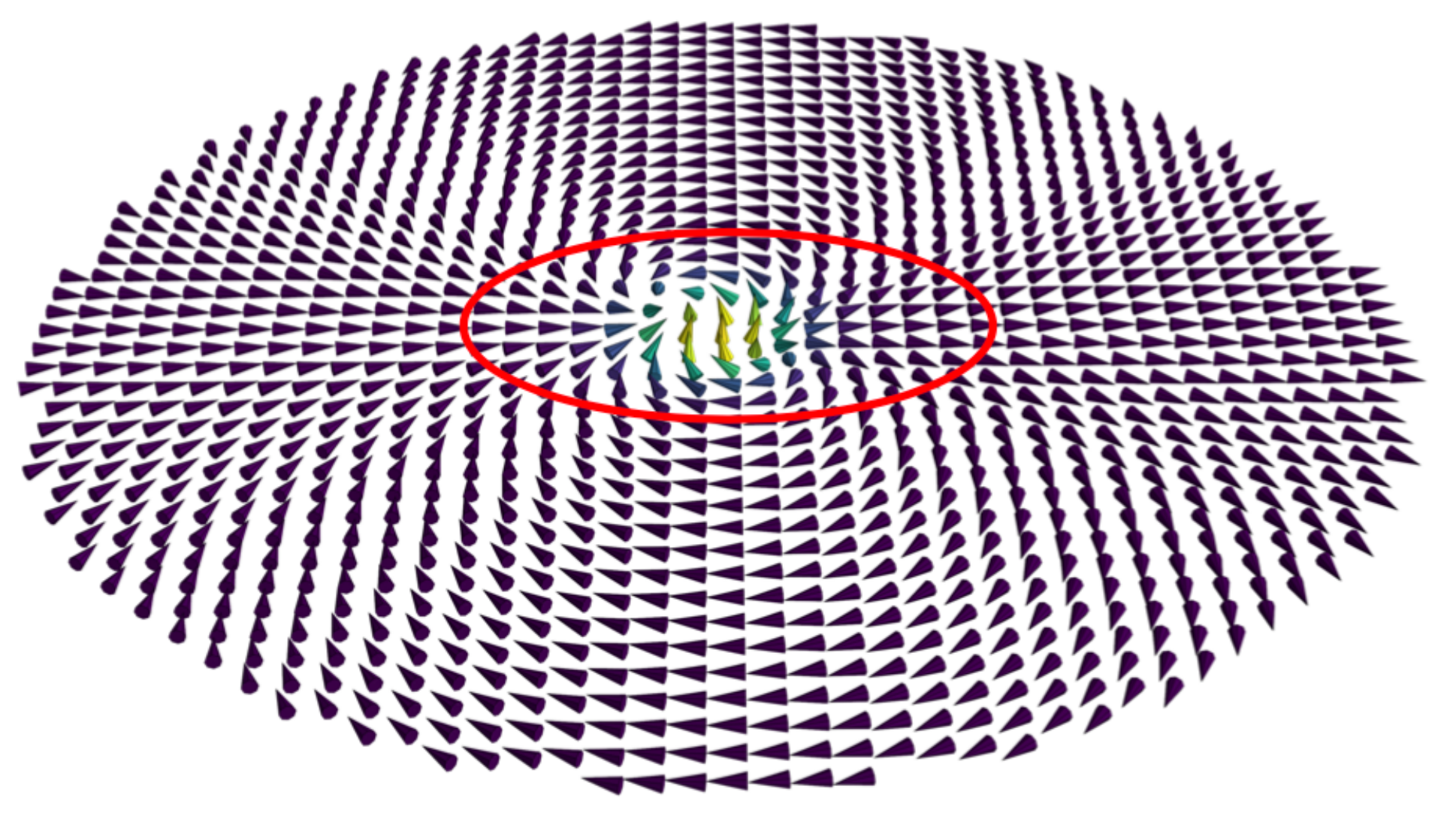}
\end{center}
\caption{A single vortex of charge $Q_V = +2$, panels (A)-(D), in $\psi_-$ component with a positive  anticlockwise circulation. Spin texture Syrmion of charge $Q_S = +2$, panels (E)-(F). The Skyrmion charge for the circle of radius $R = 5.0$ boundary curve in (E) using (\ref{QS-final}) yields $Q_2 \approx 1.961$ and $Q_1 \approx 0.036$, $Q_S = Q_2 + Q_1 \approx 1.997$. The chemical potential and the frequency trap are $\mu = 21.86869$ are $\omega^2 = 0.3$. 
The magnetic field is $\vec B_T = \mu_0 {\cal B}_0 ((x^2 - y^2) \hat x + 2 x y \hat y)$, panel (G), and $B_z = \mu_0 B_0 = 0.0$, with $\mu_0 {\cal B}_0 = 0.1$.
} \label{skyrm-2}
\end{figure}

\newpage

\newpage
\section*{References}
\bibliography{bibliography}

\end{document}